\documentclass[12pt]{article}
\usepackage{amssymb}
\usepackage{times}
\usepackage{titling}
\usepackage{graphicx}
\usepackage{amsmath}
\usepackage{color}
\usepackage{bm}
\usepackage{lineno}
\usepackage[paper=letterpaper,twoside,top=1.8cm,bottom=1.8cm,left=1.5cm,
            right=1.5cm,bindingoffset=0.0cm,voffset=0.0cm]{geometry}
\usepackage{titling}
\usepackage{multirow}

\newcommand{\dmso}{Dy$_3$Mg$_2$Sb$_3$O$_{14}$}
\newcommand{\be}{\begin{equation}}
\newcommand{\ee}{\end{equation}}
\newcommand{\bea}{\begin{eqnarray}}
\newcommand{\eea}{\end{eqnarray}}

       \renewcommand\refname{References}
        \let\oldthebibliography=\thebibliography
        \let\oldendthebibliography=\endthebibliography
        \renewenvironment{thebibliography}[1]{
            \oldthebibliography{#1}%
            \setcounter{enumiv}{0}%
        }{\oldendthebibliography}
        \renewcommand{\figurename}{Figure}

\title{ {\bf Emergent Order in the Kagome Ising Magnet Dy$_3$Mg$_2$Sb$_3$O$_{14}$} \\}
\author{
Joseph A.~M. Paddison$^{1,2,\dagger}$, Harapan S. Ong$^1$, James O. Hamp$^1$, \\ 
Paromita Mukherjee$^1$, Xiaojian Bai$^2$, Matthew G. Tucker$^{3}$, Nicholas P. Butch$^4$, \\
Claudio Castelnovo$^1$, Martin Mourigal$^2$, and S.~E. Dutton$^{1,\ast}$ \\ \\
	  	\normalsize{$^1$Cavendish Laboratory, University of Cambridge, JJ Thomson Avenue, Cambridge, CB3 0HE, UK } \\
		\normalsize{$^2$School of Physics, Georgia Institute of Technology, 	Atlanta, GA 30332, USA }\\
		\normalsize{$^3$ISIS Neutron and Muon Source, Rutherford Appleton Laboratory,} \\
		\normalsize{Harwell Campus, Didcot, OX11 0QX, UK }\\
		\normalsize{$^4$NIST Center for Neutron Research, National Institute of Standards and Technology,} \\ 
		\normalsize{Gaithersburg, MD 20899, USA} \\
		\\
		\normalsize{$^\dagger$Email: paddison@gatech.edu} \\
		\normalsize{$^\ast$Email: sed33@cam.ac.uk} \\
}
\date{April 29, 2016}

\begin{document}
\maketitle
\baselineskip22pt

\textbf{The kagome lattice---a two-dimensional (2D) arrangement of corner-sharing triangles---is at the forefront of the search for exotic states generated by magnetic frustration. Such states have been observed experimentally for Heisenberg \cite{Vries_2009,Han_2012,Fak_2012,Mendels_2016} and planar \cite{Zhou_2007,Zorko_2010,Cairns_2016} spins. In contrast, frustration of Ising spins on the kagome lattice has previously been restricted to nano-fabricated systems \cite{Mengotti_2011,Zhang_2013,Anghinolfi_2015} and spin-ice materials under applied magnetic field \cite{Fennell_2007,Matsuhira_2002a}. Here, we show that the layered Ising magnet \dmso~\cite{Dun_2016} hosts an emergent order predicted theoretically for individual kagome layers of in-plane Ising spins \cite{Moller_2009,Chern_2011}. Neutron-scattering and bulk thermomagnetic measurements, supported by Monte Carlo simulations, reveal a phase transition at $T^{\ast} \approx 0.3$\,K from a disordered spin-ice like regime \cite{Wills_2002} to an ``emergent charge ordered" state \cite{Moller_2009,Chern_2011} in which emergent charge degrees of freedom exhibit three-dimensional order while spins remain partially disordered. Our results establish \dmso~as a tuneable system to study interacting emergent charges arising from kagome Ising frustration.}

\clearpage
\baselineskip18pt

The Ising model---in which degrees of freedom (spins) are binary valued (up/down)---is a cornerstone of statistical physics that shows rich behaviour when spins occupy a highly-frustrated lattice such as kagome \cite{Wills_2002,Kim_2010,Han_2008}. If spins lie within kagome planes and point either towards or away from the centre of each triangle, the potential for emergent behaviour is shown by considering a spin (magnetic dipole) as two separated $+$ and $-$ magnetic charges: the ``emergent charge" $\cal{T}$ of a triangle \cite{Chern_2011} is defined as the algebraic sum over the three charges it contains [Fig.~1a]. Ferromagnetic nearest-neighbour interactions favour $\mathcal{T}=\pm1$ states, yielding six degenerate states on each triangle. This macroscopic ground-state degeneracy leads to a zero-point entropy, $S_\mathrm{0} \approx \frac{1}{3} R\ln \frac{9}{2}$, and suppresses spin order \cite{Wills_2002}, in analogy to 3D ``spin ice" materials \cite{Bramwell_2001,Castelnovo_2008}. The long-range magnetic dipolar interaction generates an effective Coulomb interaction between emergent charges, driving a transition to an  ``emergent charge ordered" (ECO) state entirely absent in spin ice \cite{Moller_2009,Chern_2011}. In the ECO state, $+$ and $-$ charges alternate, but the remaining threefold degeneracy of spin states for each charge means that spin order is only partial [Fig.~1b]. The ECO state has two bulk experimental signatures: nonzero entropy $S_{\mathrm{0}} \approx 0.11R$ \cite{Moller_2009}, and the presence of both Bragg and diffuse magnetic scattering in neutron-scattering measurements \cite{Brooks-Bartlett_2014}. State-of-the art experimental studies of nano-fabricated systems have allowed ECO to be inferred in the 2D limit \cite{Mengotti_2011,Zhang_2013,Anghinolfi_2015}, but a crucial experimental observation has remained elusive---namely, observation of the spatial arrangement of emergent charges in a bulk material.

\begin{figure}[th!]
\begin{center}
	\includegraphics{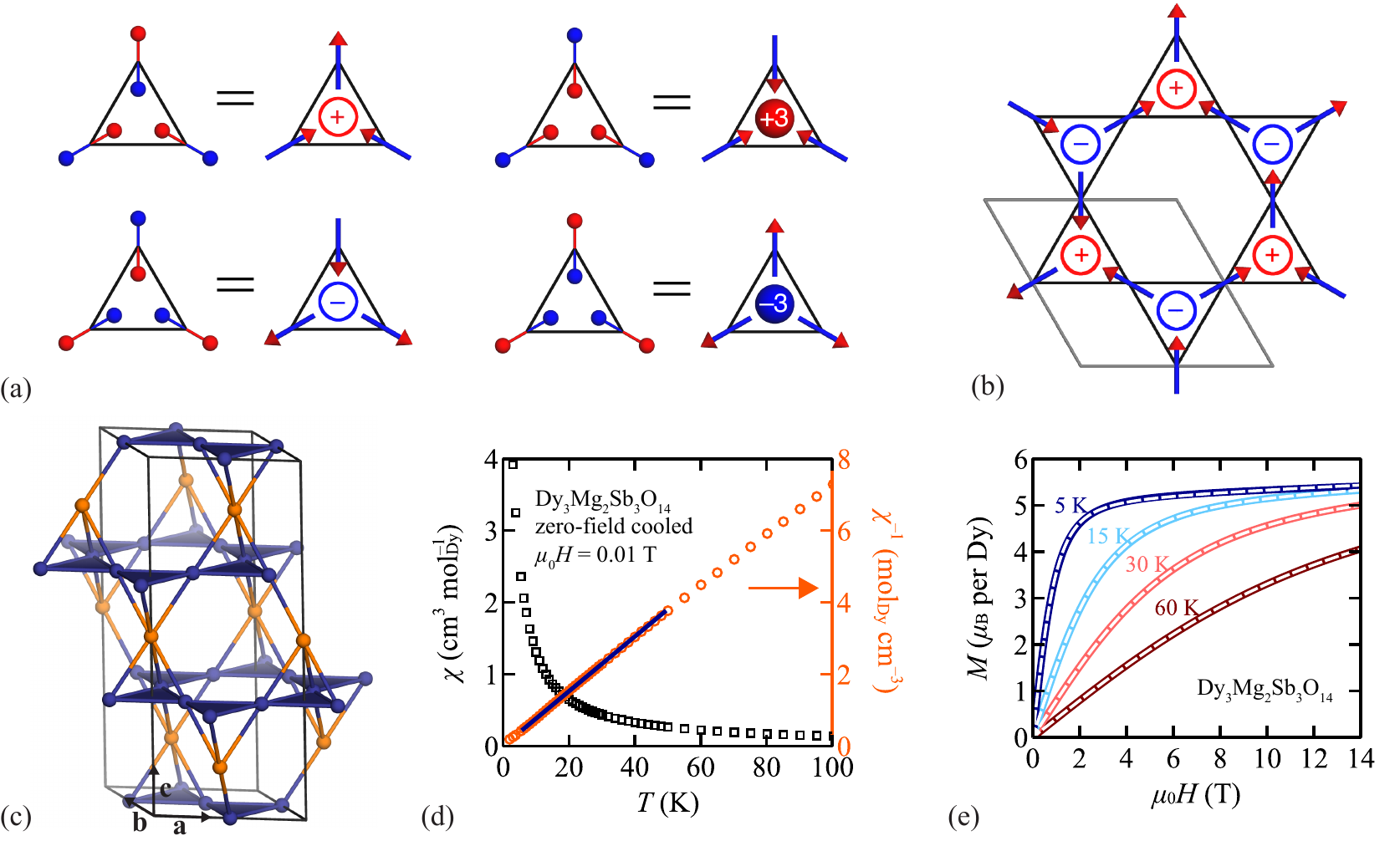} 
    \caption{\textbf{Ising spins on the kagome lattice.} (a) Relationship between spin vectors (arrows), magnetic dipoles (connected red and blue spheres), and emergent charge $\mathcal{T}$ of a triangle. (b) Example of a microstate showing emergent charge order (ECO). (c) Partial crystal structure of \dmso, showing kagome Dy$_{1-x}$Mg$_x$ site (blue spheres) and triangular Mg$_{1-3x}$Dy$_{3x}$ site (orange spheres), where $x = 0.06(2)$. (d) Magnetic susceptibility data $\chi(T)$ measured in an applied field $\mu_0 H =0.01$\,T after zero-field cooling (left axis; black squares), inverse magnetic susceptibility data $\chi^{-1}$ (right axis; orange circles), and Curie-Weiss fit over the range $5 \leq T \leq 50$\,K (blue line). (e) Dependence of magnetization $M$ on applied magnetic field $\mu_0 H$ at different temperatures (labelled above each curve) and fits to the paramagnetic Ising model. Data are shown as solid coloured lines and fits as white dashed lines (note the nearly perfect agreement: as plotted the fit lines are indistinguishable from the data).}
\end{center}
\end{figure}

Structural and magnetic characterizations suggest that the bulk magnet \dmso~is an ideal candidate for an ECO state \cite{Dun_2016}. The material crystallizes in a variant of the pyrochlore structure (space group $R\bar{3}m$) in which kagome planes of magnetic Dy$^{3+}$  alternate with triangular layers of non-magnetic Mg$^{2+}$ [Fig.~1c]. X-ray and neutron powder diffraction measurements confirm the absence of a structural phase transition to 0.2\,K [Fig.~S1] and reveal a small amount of site disorder in our sample, with 6(2)\% of Dy kagome sites occupied by Mg (and 18(6)\% of Mg sites occupied by Dy). Curie-Weiss fits to the magnetic susceptibility [Fig.~1d] indicate weak net spin interactions (the Curie-Weiss constant $\theta_\mathrm{CW}=-0.1(2)$\,K for fitting range $5\leq T\leq 50$\,K, consistent with \cite{Dun_2016}, but depends strongly on fitting range). The local environment of Dy$^{3+}$ in \dmso~is similar to the cubic spin ice Dy$_2$Ti$_2$O$_7$, suggesting that Dy$^{3+}$ spins have an Ising anisotropy axis directed ``in" or ``out" of kagome triangles with an additional component perpendicular to kagome planes. Experimentally, we confirm Ising anisotropy at low temperatures using isothermal magnetization measurements, which are ideally described by paramagnetic Ising spins with magnetic moment $\mu = 10.17(8)\,\mu_\mathrm{B}$ per Dy [Fig.~1e], and by inelastic neutron scattering [Fig.~S2].

\begin{figure}[h!]
\begin{center}
	\includegraphics{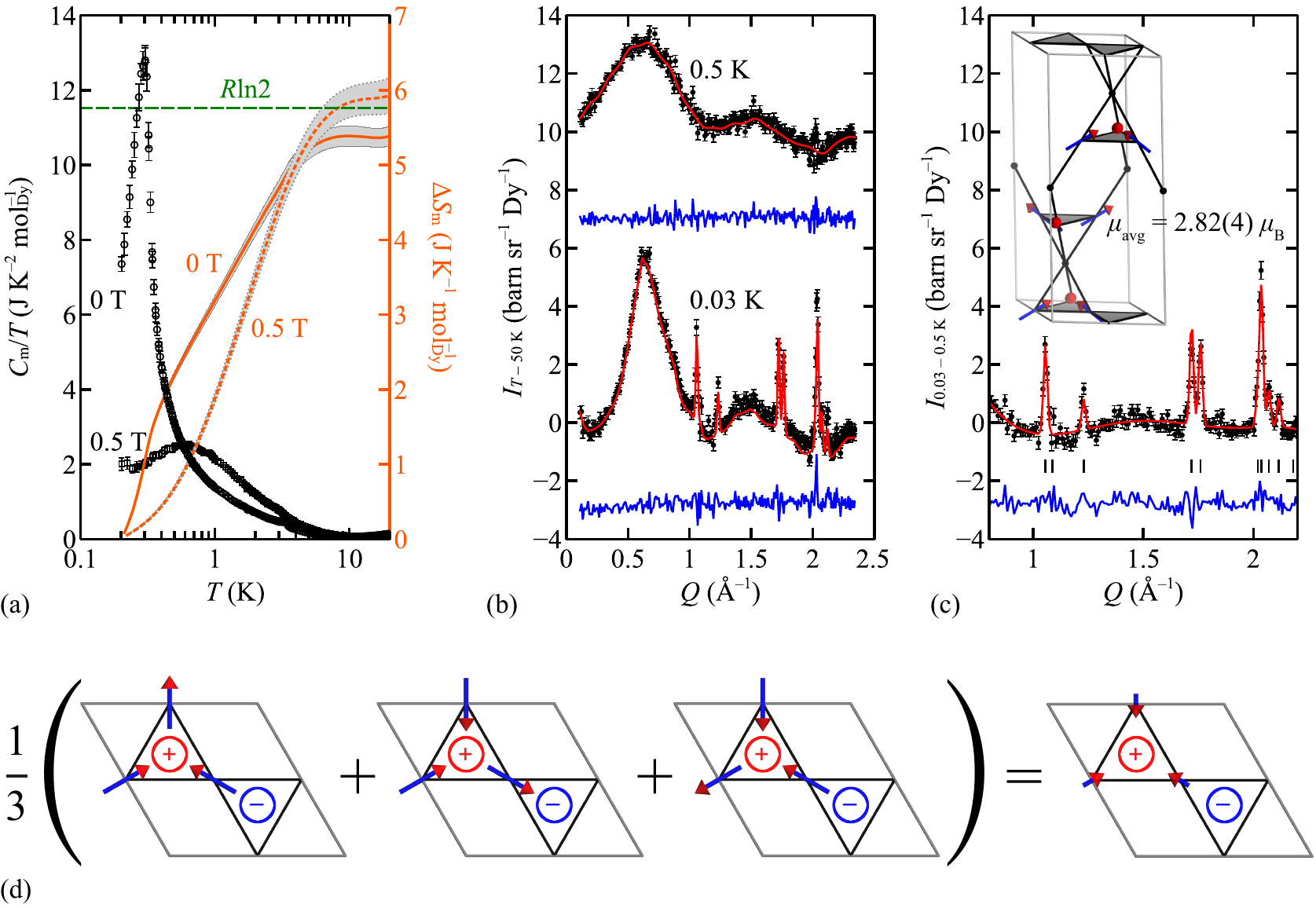} 
    \caption{\textbf{Low-temperature magnetism of \dmso.} Magnetic heat capacity divided by temperature $C_\mathrm{m}/T$ (left axis; black points) and magnetic entropy change $\Delta S_\mathrm{m}(T)$ (right axis; orange curves). Zero-field data and data measured in applied field $\mu_{0}H=0.5$\,T are shown (fields labelled on each curve). Error bars indicate one standard error throughout, unless otherwise noted. (b) Magnetic neutron-scattering data (black circles) at $T = 0.03$\,K and $0.5$\,K obtained by subtracting a high-temperature ($50$\,K) measurement as background, fits from reverse Monte Carlo (RMC) refinements (red lines), and difference (blue lines). The 0.5\,K curves are vertically shifted by 10~barn\,sr$^{-1}$\,Dy$^{-1}$ for clarity. (c) Magnetic Bragg scattering obtained as the difference between $0.03$ and $0.5$\,K data (black circles), fit from Rietveld refinement (red line), and difference (blue line). The inset shows the model of the average magnetic structure obtained from Rietveld refinement. (d) The vector average of the three microstates that are equally occupied in a ECO state yields an average ``all-in/all-out" structure with ordered moment $\mu_{\mathrm{avg}} = \mu/3$, consistent with experimental observations.}
\end{center}
\end{figure}

The magnetic specific heat $C_{\mathrm{m}}(T)$ shows that spin correlations start to develop below 5\,K and culminate in a large anomaly at $T^{\ast} = 0.31(1)$\,K that we attribute to a magnetic phase transition [Fig.~2a]. Below 0.20\,K, the spins fall out of equilibrium, as is also reported in spin-ice materials \cite{Pomaranski_2013}. In zero applied field, the entropy change $\Delta S_\mathrm{m}(T)$ from 0.2\,K to $T = 10$\,K is slightly less than the expected $R\ln2$ for random Ising spins; however, the full $R\ln2$ entropy is recovered in a small applied field of 0.5\,T. Remarkably, the $0.05(3)R$ difference between $\Delta S_\mathrm{m}(10\,\mathrm{K})$ in zero field and in a 0.5\,T field is of the same order as the entropy associated with 2D ECO ($0.11R$). Neutron-scattering experiments on a powder sample of $^{162}$\dmso~reveal the microscopic processes at play across $T^{\ast}$. Fig.~2b shows magnetic neutron-scattering data at 0.5\,K (above $T^{\ast}$) and at the nominal base temperature of 0.03\,K (below $T^{\ast}$). At 0.5\,K, our data show magnetic diffuse scattering only, which resembles spin-ice materials \cite{Hallas_2012}. In contrast, at 0.03\,K, strong magnetic diffuse scattering is observed in addition to magnetic Bragg peaks. These peaks develop at $T \leq 0.35$\,K; i.e., as $T^{\ast}$ is crossed. No additional peaks are observed on further cooling and the magnetic scattering (elastic to within the experimental resolution of 17\,$\mu$eV) does not change between 0.1\,K and 0.03\,K. Our 0.03\,K data suggest two immediate conclusions. First, the magnetic Bragg peaks are described by the propagation vector $\mathbf{k} = \mathbf{0}$; i.e., order preserves the crystallographic unit cell below $T^{\ast}$. Second, the presence of strong magnetic diffuse scattering shows that spin disorder persists below $T^{\ast}$, consistent with the observation of residual entropy at 0.2\,K and with expectations for an ECO state \cite{Brooks-Bartlett_2014}.

We use reverse Monte Carlo (RMC) refinement \cite{McGreevy_1988,Paddison_2012} to fit spin microstates to data collected between 0.03\,K and 4\,K. A single RMC microstate can capture both the average spin structure responsible for Bragg scattering and the local deviations from the average responsible for diffuse scattering [Fig.~2b]. We determine the average spin structure by two methods: first, by averaging refined RMC microstates onto a single unit cell; second, by using a combination of symmetry analysis and Rietveld refinement to model the magnetic Bragg profile (obtained as the difference between 0.03\,K and 0.5\,K data) [Fig.~2c]. Both approaches yield the same ``all-in/all-out" \emph{average} spin structure [Fig.~2c, inset]. The ordered magnetic moment at 0.03\,K, $\mu_\mathrm{avg} = 2.82(4)\,\mu_\mathrm{B}$ per Dy, is much less than the total moment of  $\mu\approx 10\,\mu_\mathrm{B}$. These results are entirely consistent with the existence of ECO. Indeed, Fig.~2d shows that averaging over the three possible ECO microstates for a given triangle generates an ``all-in/all-out" average structure, as observed experimentally; moreover, the expected ordered moment for ECO,  $\mu/3 \approx 3.3\,\mu_\mathrm{B}$ per Dy \cite{Brooks-Bartlett_2014}, is in general agreement with the measured value of $2.82(4)\,\mu_\mathrm{B}$ per Dy.

\begin{figure}[h!]
\begin{center}
	\includegraphics{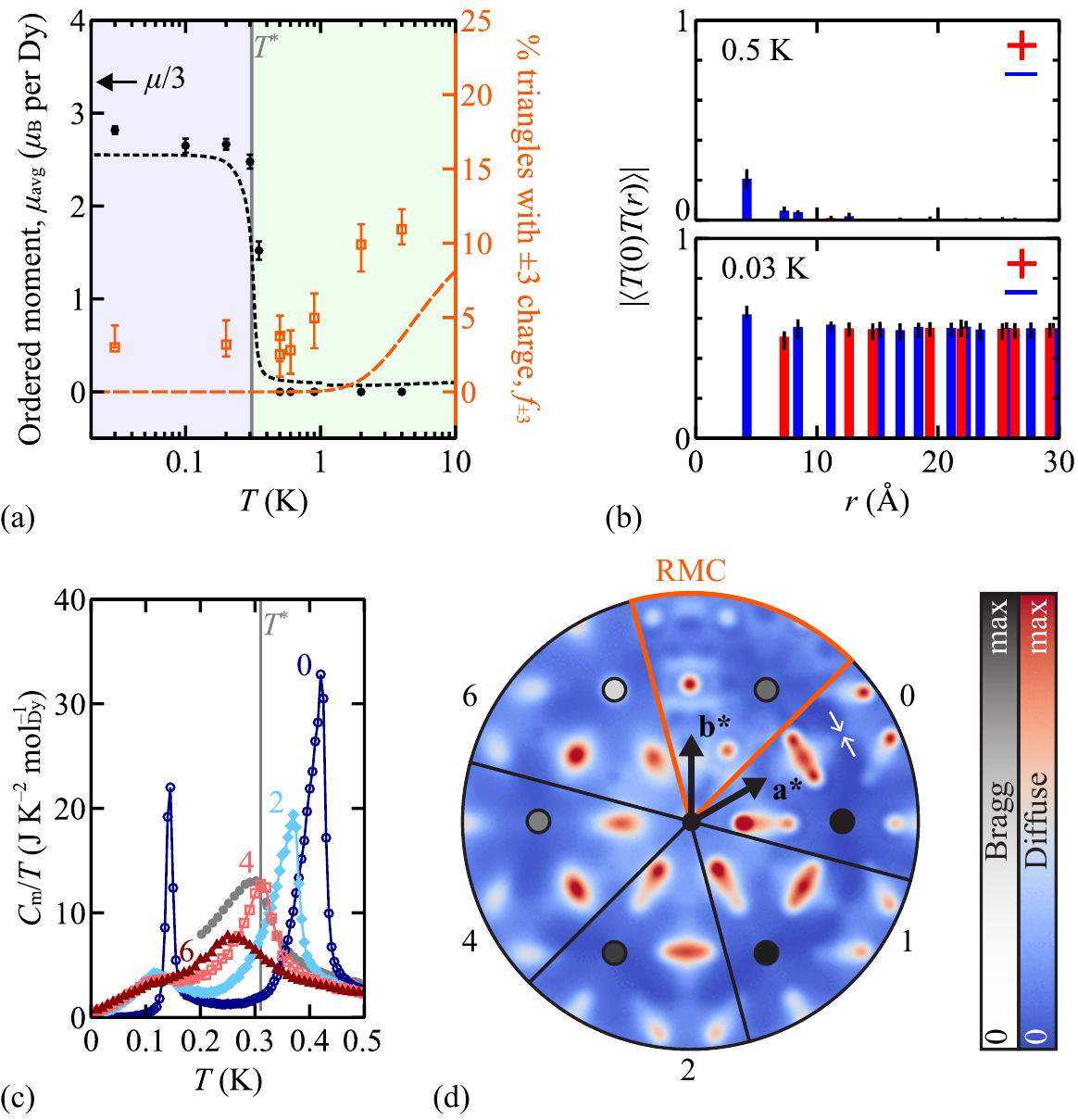} 
    \caption{\textbf{Charge-order transition, site disorder, and Coulomb phase in \dmso.}~(a) Temperature evolution of the ordered magnetic moment per Dy, $\mu_{\mathrm{avg}}$ (left axis), and the number of triangles for which $\mathcal{T}=\pm3$, $f_{\pm3}$ (right axis). Values of $\mu_{\mathrm{avg}}$ from Rietveld refinements to experimental data are shown as filled black circles, and values from MC simulations (with 4\% Mg on the Dy site) as a black dotted line. Upper bounds on $f_{\pm3}$ from RMC refinements to experimental data are shown as hollow orange squares, and values from MC simulations as an orange dashed line. The location of $T^{\ast}$ is shown by a vertical grey line, and the background is shaded blue below $T^{\ast}$ and green above $T^{\ast}$. (b) Charge-correlation function $\langle \mathcal{T}(0)\mathcal{T}(r_{ab})\rangle$ obtained from RMC refinements at $0.5$\,K (upper panel) and $0.03$\,K (lower panel). Solid bars show correlation magnitudes, with positive correlations shown in red and negative correlations in blue. Error bars on $f_{\pm3}$ and $\langle \mathcal{T}(0)\mathcal{T}(r_{ab})\rangle$ are derived assuming 10\% uncertainty on the absolute intensity normalization of the magnetic scattering data. (c) Magnetic heat capacity from MC simulations (system size $N = 7776$ spins) for different amounts of random site disorder (the \% Mg on the Dy site is labelled above each curve). (d) Single-crystal neutron scattering calculations in the $(hk0)$ plane from MC simulations at $T = 0.2$\,K for different amounts of random site disorder (the \% Mg on the Dy site is labelled on each segment of the plot). The single-crystal calculation from RMC refinement to 0.03\,K powder data (for $6\%$ Mg on the Dy site) is shown for comparison. Bragg and diffuse scattering intensities are shown using separate colour scales, and the location of a pinch point is indicated by small white arrows.}
\end{center}
\end{figure}

To look for signatures of ECO in real space, we compare the temperature evolution of $\mu_\mathrm{avg}$ with the percentage of $\mathcal{T}=\pm3$ charges [Fig.~3a]. The latter quantity, $f_{\pm3}$, takes a value of 25\% for random spins, 100\% for an ``all-in/all-out" microstate, and 0\% for a microstate that fully obeys the $\mathcal{T}=\pm1$ rule. The value of $f_{\pm3}$ extracted from RMC refinements decreases with lowering temperature to a minimum value of $<$\,$5$\% below 1\,K; these values represent upper bounds because RMC refinements were initialized from random microstates. Crucially,  below $T^{\ast}$, the $\mathcal{T}=\pm1$ rule is obeyed while $\mu_\mathrm{ord}$ is non-zero, as is required for ECO [Fig.~3a]. We further confirm ECO by calculating the charge-correlation function $\langle \mathcal{T}(0)\mathcal{T}(r_{ab})\rangle$, the average product of charges separated by radial distance $r_{xy}$  on the honeycomb lattice formed by the triangle midpoints [Fig.~3b]. At 0.5\,K, this function decays with increasing $r_{xy}$, indicating that $\mathcal{T}=\pm1$ charges are disordered. At 0.03\,K, $\langle \mathcal{T}(0)\mathcal{T}(r_{ab})\rangle$ shows two key features that indicate an ECO state: a diverging correlation length, and an alternation in sign with a negative peak at the nearest-neighbour distance [Fig.~3b]. The magnitude of $\langle \mathcal{T}(0)\mathcal{T}(r_{ab}) \rangle$ found experimentally ($\approx 0.6 = (0.94\times3\mu_{\mathrm{avg}}/\mu)^2$) is smaller than the value of unity corresponding to an ideal ECO state, which indicates that the alternation of charges contains some errors; we show below this is probably due to the presence of site disorder.

Why does \dmso~show fundamentally the same ECO as predicted for a 2D kagome system of in-plane Ising spins? This is far from obvious, because the real material differs from the existing model in three respects: i) the spins are canted 26(2)$^{\circ}$ to the kagome planes, ii) the planes are layered in 3D, and iii) there is Dy/Mg site disorder [Fig.~2c]. This puzzle is elucidated by Monte Carlo simulations for a minimal model containing the nearest-neighbour exchange interaction $J = -3.72$\,K determined for structurally-related Dy$_2$Ti$_2$O$_7$ \cite{Hertog_2000}, and the long-range magnetic dipolar interaction $D = 1.28$\,K calculated from experimentally determined Dy--Dy distances. In 2D, spin canting interpolates between two limits---an ECO transition followed by lower-temperature spin-ordering (SO) for in-plane spins \cite{Chern_2011}, and a single SO transition for spins perpendicular to kagome planes \cite{Chioar_2016}---and hence destabilises ECO compared to the 2D in-plane limit. In contrast, the stacking of kagome planes stabilizes 3D ECO---uniquely minimizing the effective Coulomb interaction between emergent charges---but leaves the SO transition temperature essentially unchanged. The effect of random site disorder is shown in Fig.~3c. Disorder broadens the specific-heat anomalies and suppresses the ECO transition temperature. In spite of this, we find that a distinct ECO phase persists for 6\% Mg on the Dy site; i.e., the estimated level of disorder present in our sample of \dmso. Moreover, simulated magnetic specific-heat [Fig.~3c] and powder neutron-scattering [Fig.~S8] curves with $\sim$$4$ to $6$\% Mg on the Dy site show remarkably good agreement with experimental data, especially given that $J$ is not optimised for \dmso.

An ECO microstate can be coarse-grained into a magnetization field with two components: the ``all-in/all-out" average spin structure with nonzero divergence, and the local fluctuations from the average that are captured by (divergence-free) dimer configurations on the dual honeycomb lattice \cite{Brooks-Bartlett_2014}. Without site disorder, the latter component yields ``pinch-point" features in single-crystal diffuse-scattering patterns, the signature of a Coulomb phase \cite{Brooks-Bartlett_2014,Fennell_2009}. Fig.~3d shows that the introduction of site disorder blurs the pinch points and reduces the magnitude of the ordered moment in the ECO phase. We find good overall agreement between patterns from model simulations with $\sim$$4$ to $6$\% Mg on the Dy site and from RMC microstates refined to powder data [Fig.~3d]. These results suggest that pinch-point scattering could be observed in single-crystal samples of \dmso~with low levels of disorder. Our simulations also suggest why a transition from ECO to SO is not observed experimentally: single-spin-flip dynamics (arguably more appropriate to real materials) become frozen in the ECO state and non-local (loop) dynamics are required to observe the SO transition in MC simulations.

The ECO state in \dmso~is the first realisation of ordering of emergent degrees of freedom in a solid-state kagome material. Phase transitions driven by emergent excitations are rare---related examples being the critical end-point in spin ice \cite{Castelnovo_2008,Sakakibara_2003,Hamp_2015} and the recently-reported fractionalized phase in Nd$_2$Zr$_2$O$_7$ \cite{Petit_2016}. Moreover, the unusually slow spin dynamics characteristic of lanthanide pyrochlore oxides offer the exciting possibility of measuring finite-time (Kibble-Zurek) scaling at the ECO critical point \cite{Hamp_2015}. Whether the predicted SO \cite{Chern_2011} eventually occurs in \dmso~remains to be seen: spin freezing \cite{Snyder_2004,Matsuhira_2011} or site disorder may prevent its onset. We expect physical and/or chemical perturbations to control the properties of \dmso; e.g., application of magnetic field slightly tilted from the $c$-axis should drive a Kastelyn transition towards SO \cite{Fennell_2007,Matsuhira_2002a}; modified synthesis conditions may allow the degree of site mixing to be controlled \cite{Dun_2016}; and application of chemical pressure may alter the spin-canting angle and/or the distance between kagome layers, potentially generating a novel SO phase instead of ECO for sufficiently large canting \cite{Chioar_2016}. Substitution of Dy$^{3+}$ by other lanthanide ions \cite{Dun_2016,Sanders_2016,Sanders_2016a,Scheie_2016} may increase the ratio of exchange to dipolar interactions, offering promising routes towards exotic spin-liquid behaviour: dimensionality reduction by effective layer decoupling (when exchange dominates over dipolar interactions), and realisation of quantum kagome systems with local spin anisotropies.

\subsection*{References}
\begingroup
\renewcommand{\section}[2]{}

\endgroup


\subsection*{Methods}

\noindent\textit{Sample preparation.} Powder samples of \dmso~were prepared from a stoichiometric mixture of dysprosium (III) oxide (99.99\%, Alfa Aesar \cite{NIST}), magnesium oxide (99.998\%, Alfa Aesar \cite{NIST}), and antimony (V) oxide (99.998\%, Alfa Aesar \cite{NIST}). For neutron-scattering experiments a $\sim$5\,g sample isotopically enriched with $^{162}$Dy (94.4(2)\% $^{162}$Dy$_2$O$_3$, CK Isotopes~\cite{NIST}) was prepared. For all samples, starting materials were intimately mixed and pressed into pellets before heating at 1350\,$^{\circ}$C for 24 hours in air. This heating step was repeated until the amount of impurity phases as determined by X-ray diffraction was no longer reduced on heating. The enriched sample contained impurity phases of MgSb$_2$O$_6$ (6.4(5)\,wt\%) and Dy$_3$SbO$_7$ (0.97(8)\,wt\%), the latter of which orders antiferromagnetically at $T\approx3$\,K \cite{Fennell_2001}. \\

\noindent\textit{X-ray diffraction measurements.} Powder X-ray diffraction (XRD) was carried out using a Panalytical Empyrean diffractometer \cite{NIST} with Cu $K\alpha$ radiation ($\lambda = 1.5418 $\,\AA). Measurements were taken between $5\leq 2\theta \leq 120^\circ$ with $\Delta2\theta = 0.02^\circ$.  \\

\noindent\textit{Neutron scattering measurements.} Powder neutron diffraction measurements were carried out on the GEM diffractometer at the ISIS Neutron and Muon Source, Harwell, UK \cite{Hannon_2005}, at $T =0.50,0.60,0.90,2.0,4.0,25$, and 300\,K. For $T=25$ and 300\,K measurements, around 4.2\,g of isotopically-enriched powder was loaded into a $\phi = 6$\,mm vanadium can and cooled in a flow cryostat. For measurements at $T\leq25$\,K, the same sample was loaded into a $\phi = 6$\,mm vanadium can, which was attached directly to a dilution refrigerator probe and loaded within a flow cryostat. Inelastic neutron-scattering experiments were carried out on the Disk Chopper Spectrometer (DCS) at the NIST Center for Neutron Research, Gaithersburg MD, USA \cite{Copley_2003}, at $T=0.03,0.10,0.20,0.30,0.35$, and $0.50$\,K. Around $1.1$\,g of isotopically-enriched powder was loaded into a $\phi = 4.7$\,mm copper can and mounted at the base of a dilution refrigerator. The temperature was measured at the mixing chamber and does not necessarily reflect the sample temperature for 0.1\,K and 0.03\,K, as the spins progressively fall out of equilibrium. On DCS, the incident neutron wavelength was $5$\,\AA~and data were integrated over the energy range $-0.15 \leq E \leq 0.15$\,meV to obtain the total scattering [Fig.~S4]. Data reduction was performed using the MANTID and DAVE \cite{Azuah_2009} programs. All data were corrected for detector efficiency using a vanadium standard, normalized to beam current (GEM) or incident beam monitor (DCS), and corrected for absorption by the sample.\\

\noindent\textit{Crystal structure refinements.} Combined Rietveld analysis of the 300\,K X-ray and neutron (GEM) diffraction data was carried out using the FULLPROF suite of programs \cite{Rodriguez-Carvajal_1993a}. The individual patterns were weighted so that the total contribution from X-ray and neutron diffraction was equal; i.e., data from each of the five detector banks on GEM was assigned 20\% of the weighting of the single X-ray pattern. The neutron scattering cross-section for Dy was fixed to $b_\mathrm{Dy} = -0.6040$\,fm, to reflect the isotopic composition as determined by inductively coupled plasma mass spectrometry (ICP-MS). Peak shapes were modelled using a pseudo-Voigt function, convoluted with an Ikeda-Carpenter function or an axial divergence asymmetry function for neutron and X-ray data respectively. Backgrounds were fitted using a Chebyshev polynomial function. At 25\,K, Rietveld analysis of only the neutron diffraction data was carried out. In addition to the impurity phases observed in X-ray diffraction, a small amount ($<$\,$1$\,wt\%) of vanadium (IV) oxide from corrosion of the vanadium sample can was also observed in the neutron-diffraction data. The fit to 300\,K data is shown in Fig.~S1, refined values of structural parameters are given in Table~S1, and selected bond lengths are given in Table~S2.\\

\noindent\textit{Magnetic measurements.} Magnetic susceptibility measurements, $\chi(T)=M(T)/H$, were made using a Quantum Design Magnetic Properties Measurement System (MPMS~\cite{NIST}) with a superconducting interference device (SQUID) magnetometer. Measurements were made after cooling in zero field (ZFC) and in the measuring field (FC) of $\mu_{0}H=0.1$\,T over the temperature range $2 \leq T \leq 300$\,K. Isothermal magnetization $M(H)$ measurements were made using a Quantum Design Physical Properties Measurement System (PPMS~\cite{NIST}) at selected temperatures $1.6 \leq T \leq 80$\,K between $-14 \leq \mu_{0}H \leq 14$\,T. A global fit to the $M(H)$ data for $T \geq 5$\,K [Fig.~1e] was performed using the powder-averaged form for free Ising spins,
\begin{equation}
M_{\mathrm{Ising}}^{\mathrm{powder}}=\frac{\mu}{2}\int_{-1}^{1}\cos\theta\tanh\left(\frac{\mu H\cos\theta}{k_{\mathrm{B}}T}\right)\mathrm{d}(\cos\theta),
\end{equation}
where $H$ is applied magnetic field, and magnetic moment $\mu$ is the only fitting parameter \cite{Bramwell_2000}. The fitted value $\mu = 10.17(8)\,\mu_\mathrm{B}$ per Dy is in close agreement with the expected value of $10.0\,\mu_\mathrm{B}$ for a Kramers doublet ground state with $g = 4/3$ and $m_J = \pm15/2$; in particular, the reduced value of the saturated magnetization, $M_\mathrm{sat} \approx \mu/2$, is as expected for powder-averaged Ising spins \cite{Bramwell_2000}. \\

\noindent\textit{Heat capacity measurements.} Heat capacity measurements were carried out on a Quantum Design PPMS~\cite{NIST} instrument using dilution fridge ($0.07 \leq T\leq 4$\,K) and standard ($1.6 \leq T \leq 250$\,K) probes in a range of measuring fields, $0 \leq \mu_0 H \leq 0.5$\,T. To ensure sample thermalisation at low temperatures, measurements were made on pellets of \dmso~mixed with an equal mass of silver powder, the contribution of which was measured separately and subtracted to obtain $C_p$. The magnetic specific heat $C_\mathrm{m}$ was obtained by subtracting modelled lattice $C_\mathrm{l}$ and nuclear $C_\mathrm{n}$ contributions from $C_p$. We obtained $C_\mathrm{l}$ by fitting an empirical Debye model to the $10 < T< 200$\,K data, with $\theta_\mathrm{D} =272(13)$\,K. To obtain a lower bound on the contact hyperfine and electronic quadrupolar contributions to $C_p$ \cite{Pomaranski_2013,Henelius_2016}, we used previous experimental results on dysprosium gallium garnet (DGG) \cite{Filippi_1977}, a related material for which these contributions are known down to $T = 0.037$\,K. Correcting for the larger static electronic moment $\approx4.2\,\mu_\mathrm{B}$ of DGG compared to $\langle \mu \rangle \geq 2.5\,\mu_\mathrm{B}$ below 0.2\,K for \dmso, we obtained the high-temperature tail of the nuclear hyperfine contributions as $C_p = A/T^2$ with $A = 0.0032$\,J\,K\,mol$_{\mathrm{Dy}}^{-1}$ [Fig.~S3]. \\

\noindent\textit{Magnetic total scattering.} To isolate the magnetic contribution to the neutron-scattering data, data collected at a high temperature $T_\mathrm{high}\gg\theta_\mathrm{CW}$ was subtracted from the low-temperature data of interest, where $T_\mathrm{high}=25$\,K (GEM data) or 50\,K (DCS data). For the data obtained below the magnetic ordering temperature of the Dy$_3$SbO$_7$ impurity phase ($\approx3$\,K \cite{Fennell_2001}), a refined model of the magnetic Bragg scattering of Dy$_3$SbO$_7$ was subtracted, as described in Section~S5; we note that the orthorhombic crystal structure of Dy$_3$SbO$_7$ \cite{Siqueira_2013} allowed the impurity Bragg peaks to be readily distinguished from sample peaks. The data were placed on an absolute intensity scale (barn\,sr$^{-1}$\,Dy$^{-1}$) by normalization to the calculated nuclear Bragg profile at $T_\mathrm{high}$. \\

\noindent\textit{Average magnetic structure analysis.} Magnetic refinements to the Bragg profile for the was carried out using the Rietveld method within the FULLPROF suite of programs \cite{Rodriguez-Carvajal_1993a}, as described above. For the magnetic-structure refinement shown in Fig.~2c, candidate magnetic structures were determined using symmetry analysis \cite{Wills_2001} via the SARAH \cite{Wills_2000} and ISODISTORT \cite{Campbell_2006} programs. The average magnetic structure is described by the irreducible representation $\Gamma_3$, in Kovalev's notation \cite{Kovalev_1993}. The basis vectors of the magnetic structure are given in Table~S5 and refined values of structural parameters are given in Table~S6. \\

\noindent\textit{Reverse Monte Carlo refinements.} Refinements to the total (Bragg$+$diffuse) magnetic scattering were performed using a modified version of the SPINVERT program \cite{Paddison_2013}. In these refinements, a microstate was generated as a periodic supercell containing $N=7776$ Dy$^{3+}$ spin vectors $\mathbf{S}_i = \mu\sigma_i\mathbf{\hat{e}}_i$, where $\mu = 10.0\,\mu_\mathrm{B}$ is the fixed magnetic moment length, the unit vector  $\mathbf{\hat{e}}_i$ specifies the local Ising axis determined from Rietveld refinement, and the Ising variable $\sigma_i = \pm 1$. A random site-disorder model with 6\% non-magnetic Mg on the Dy site was assumed, and $\mathbf{S}_i\equiv \mathbf{0}$ for atomic positions occupied by Mg. Ising variables were initially assigned at random, and then refined against experimental data in order to minimise the sum of squared residuals,
\begin{equation}
\chi^{2}=W\sum_{Q}\left[\frac{I_{\mathrm{calc}}(Q)-I_{\mathrm{expt}}(Q)}{\sigma(Q)}\right]^{2},\label{eq:chi_sq}
\end{equation}
where $I(Q)$ is the magnetic total-scattering intensity at $Q$, subscripts ``calc'' and ``expt'' denote calculated and experimental intensities, respectively, $\sigma(Q)$ is an experimental uncertainty, and $W$ is an empirical weighting factor. For data collected on GEM, a refined flat-in-$Q$ background term was included in the calculated $I(Q)$. For data collected at $T\leq0.35$\,K, we obtain $I_{\mathrm{calc}}(Q)=I_{\mathrm{Bragg}}(Q)+I_{\mathrm{diffuse}}(Q)-I_\mathrm{random}(Q)$, where subscripts ``Bragg", ``diffuse", and ``random" indicate magnetic Bragg, magnetic diffuse, and high-temperature contributions, respectively. Here, $I_\mathrm{random}(Q)=\frac{2}{3}C[\mu f(Q)/\mu_\mathrm{B}]^2$, where the constant $C=(\gamma_{\mathrm{n}}r_{\mathrm{e}}/2)^2=0.07265$\,barn and $f(Q)$ is the Dy$^{3+}$ magnetic form factor \cite{Brown_2004}. The Bragg and diffuse contributions were separated by applying the identity $\mathbf{S}_{i}\equiv \langle \mathbf{S}_{i}\rangle +\Delta\mathbf{S}_i$ to each atomic position \cite{Frey_1995}, where the average spin direction $\langle \mathbf{S}_{i}\rangle$ is obtained by vector averaging the supercell onto a single unit cell, and the local spin fluctuation $\Delta\mathbf{S}_i \equiv\mathbf{S}_{i}-\langle \mathbf{S}_{i}\rangle $. The Bragg contribution is given by
\begin{equation}
I_{\mathrm{Bragg}}(Q)=C\left[\frac{f(Q)}{\mu_\mathrm{B}}\right]^2\frac{2\pi^2 N_{\mathrm{c}}}{NV}\sum_{\mathbf{G}}\frac{\left|\mathbf{F}^{\perp}(\mathbf{G})\right|^{2}}{G^2}R(Q-G),\label{eq:Bragg_powder}
\end{equation}
in which $\mathbf{G}$ is a reciprocal lattice vector with length $G$, $V$ is the volume of the unit cell, $N_\mathrm{c}$ is number of unit cells in the supercell, $R(Q-G)$ is the resolution function determined from Rietveld refinement \cite{Mellergard_1999}. The magnetic structure factor $\mathbf{F}^{\perp}(\mathbf{G})=\sum_i \left\langle \mathbf{S}_{i}\right\rangle ^{\bot}\exp\left(\mathrm{i}\mathbf{G}\cdot\mathbf{r}_{i}\right)$, where supercript ``$\perp$" indicates projection perpendicular to $\mathbf{G}$, and the sum runs over all atomic positions in the unit cell. The diffuse contribution is given by
\begin{equation}
I_\mathrm{diffuse}(Q)=C\left[\frac{f(Q)}{\mu_\mathrm{B}}\right]^2 \frac{1}{N}\left\{ \frac{2}{3}\sum_{i}|\Delta\mathbf{S}_{i}|^{2}+\sum_{j\neq i}\left[A_{ij}\frac{\sin Qr_{ij}}{Qr_{ij}}+B_{ij}\left(\frac{\sin Qr_{ij}}{\left(Qr_{ij}\right)^{3}}-\frac{\cos Qr_{ij}}{\left(Qr_{ij}\right)^{2}}\right)\right]\right\} ,\label{eq:theory_powder-1}
\end{equation}
where sums run over all atomic positions in the supercell, $r_{ij}$ is the radial distance between positions $i$ and $j$, and the correlation coefficients
$A_{ij} = \Delta\mathbf{S}_{i}\cdot\Delta\mathbf{S}_{j}-(\Delta\mathbf{S}_{i}\cdot\mathbf{r}_{ij})(\Delta\mathbf{S}_{j}\cdot\mathbf{r}_{ij})/r_{ij}^2$ and $B_{ij} = 3(\Delta\mathbf{S}_{i}\cdot\mathbf{r}_{ij})(\Delta\mathbf{S}_{j}\cdot\mathbf{r}_{ij})/r_{ij}^2-\Delta\mathbf{S}_{i}\cdot\Delta\mathbf{S}_{j}$  \cite{Blech_1964,Paddison_2013}. For data collected at $T\geq0.5$\,K, which show no magnetic Bragg scattering, we obtain $I_{\mathrm{calc}}(Q)=I_{\mathrm{diffuse}}(Q)-I_\mathrm{random}(Q)$, where $\mathbf{S}_i$ replaces $\Delta\mathbf{S}_{i}$ everywhere. All refinements employed the Metropolis algorithm with single-spin flip dynamics, and were performed for 200 proposed flips per spin, after which no significant reduction in $\chi^2$ was observed. Fits-to-data at $T = 0.03, 0.20, 0.50, 0.60, 0.90, 2.0$, and $4.0$\,K are shown in Fig.~S7.
\\

\noindent\textit{Monte Carlo simulations.} Simulations were performed for the dipolar spin ice model \cite{Hertog_2000,Melko_2004}, extended to the geometry of interest in this work. The model is defined for Ising spins $\mathbf{S}_i = \mu\sigma_i \mathbf{\hat{e}}_i$, which are constrained to point along the local easy-axis directions $\mathbf{\hat{e}}_i$ and can thus be described by the Ising pseudospin variables, $\sigma_i = \pm 1$. The Hamiltonian comprises an exchange term of strength $J$ between nearest-neighbour spins $\langle i,j \rangle$, and long-range dipolar interactions of characteristic strength $D=(\mu_0/4\pi) \mu^2 /r_{\mathrm{nn}}^3$ between all pairs of spins, where  $\mu \approx 10\,\mu_\mathrm{B}$ is the magnitude of the Dy$^{3+}$ spin and $r_\mathrm{nn}$ is the nearest-neighbour distance of the lattice. The Hamiltonian is thus given by
\begin{align}
	\mathcal{H}  = &   - J \sum_{\langle i,j \rangle} \sigma_i \sigma_j (\hat{\mathbf{e}}_i \cdot \hat{\mathbf{e}}_j)+ D r_\mathrm{nn}^3 \sum_{i>j} \sigma_i \sigma_j \left( \frac{\hat{\mathbf{e}}_i \cdot \hat{\mathbf{e}}_j}{{r}_{ij}^3} 
- \frac{3 (\hat{\mathbf{e}}_i \cdot \mathbf{r}_{ij})(\hat{\mathbf{e}}_j \cdot \mathbf{r}_{ij})}{{r}_{ij}^5} \right),
\end{align}
where $\mathbf{r}_{ij}$ is the vector of length $r_{ij}$ connecting spins $i$ and $j$. We use $D = 1.28$\,K as calculated from experimentally determined Dy--Dy distances and $J = -3.72$\,K from Dy$_2$Ti$_2$O$_7$ \cite{Hertog_2000}. We treat the long-range dipolar interactions using Ewald summation \cite{Melko_2004,Leeuw_1980} with tinfoil boundary conditions at infinity. In simulations including site disorder, non-magnetic ions are simulated by setting the corresponding $\sigma_i$ to zero. Our unit cell comprises three stacked kagome layers, each layer made from four kagome triangles. The whole system comprises $N=7776$ spins in total, commensurate with the possible $\sqrt{3}\times \sqrt{3}$ spin-ordered state found in 2D \cite{Chern_2011}. We use both single-spin flip and loop dynamics \cite{Melko_2004, Melko_2001}, with Metropolis weights. Loop dynamics are necessary to ensure ergodicity at low temperatures and explore possible long-range spin-ordered states. We use the short loop algorithm \cite{Melko_2004,Melko_2001}. One MC sweep is defined as $N$ single spin-flip attempts, followed by the proposal of loop moves until the cumulative number of proposed spin-flips (in the loops) is at least $N$. We use an annealing protocol, initializing the system at high temperature with $\sim$\,$10^{4}N$ single spin-flip attempts, then decrease the temperature incrementally. After each temperature decrement, the system is updated with $\sim$\,$10^3$ MC sweeps to ensure equilibration before collecting data every $\sim$\,$10$ MC sweeps. Powder-averaged magnetic neutron-scattering patterns calculated from MC are shown in Fig.~S8.

\subsection*{Methods References}

\begingroup
\renewcommand{\section}[2]{}

\endgroup


\subsection*{Acknowledgements}
Work at Cambridge was supported through the Winton Programme for the Physics of Sustainability. The work of J.A.M.P., X.B., and M.M. and facilities at Georgia Tech were supported by the College of Sciences through M.M. start-up funds. H.S.O. acknowledges a Teaching Scholarship (Overseas) from the Ministry of Education, Singapore. J.O.H. is grateful to the Engineering and Physical Sciences Research Council (EPSRC) for funding. C.C. was supported by EPSRC Grant No. EP/G049394/1, and the EPSRC NetworkPlus on ``Emergence and Physics far from Equilibrium". Experiments at the ISIS Pulsed Neutron and Muon Source were supported by a beamtime allocation from the Science and Technology Facilities Council. This work utilized facilities at the NIST Center for Neutron Research.  MC simulations were performed using the Darwin Supercomputer of the University of Cambridge High Performance Computing Service (http://www.hpc.cam.ac.uk/) and the ARCHER UK National Supercomputing Service (http://www.archer.ac.uk/, for which access was provided by an ARCHER Instant Access scheme). We thank  G.-W. Chern, J. Goff, A. L. Goodwin, G. Lonzarich,  G. Moller, D. Prabhakaran, J. R. Stewart, and A. Zangwill for valuable discussions, and M. Kwasigroch for preliminary theoretical work. 

\subsection*{Author contributions}
H.S.O., P.M., and S.E.D prepared the samples. H.S.O., P.M., X.B., M.M., and S.E.D. performed and analyzed the thermo-magnetic measurements. J.A.M.P., P.M., X.B., M.G.T., N.P.B., and S.E.D. performed the neutron-scattering measurements and J.A.M.P., M.M., and S.E.D. analyzed the data. J.A.M.P. carried out the RMC refinements.  J.O.H. and C.C. carried out the MC simulations. C.C. and S.E.D. conceived the project, which was supervised by C.C., M.M. and S.E.D.. J.A.M.P. wrote the paper with input from all authors.

\subsection*{Additional information}
Correspondence and requests for materials should be addressed to J.A.M.P. (paddison@gatech.edu) and S.E.D. (sed33@cam.ac.uk).

\subsection*{Competing financial interests}
The authors declare no competing financial interests.

\subsection*{Data accessibility}
The underlying research materials can be accessed at the following location: http://dx.doi.org/tbd.

\clearpage


       \renewcommand\refname{References}
       \renewcommand{\thesection}{S.\arabic{section}}
       \renewcommand{\thesubsection}{\thesection.\arabic{subsection}}
        \setcounter{equation}{0}
        \makeatletter 
        \def\tagform@#1{\maketag@@@{(S\ignorespaces#1\unskip\@@italiccorr)}}
        \makeatother
        \setcounter{figure}{0}
        \makeatletter
        \makeatletter \renewcommand{\fnum@figure}
        {\figurename~S\thefigure}
        \makeatother
        \setcounter{table}{0}
        \makeatletter
        \makeatletter \renewcommand{\fnum@table}
        {\tablename~S\thetable}
        \makeatother

\title{ {Supplementary Information} \\[10px]
{\bf Emergent Order in the Kagome Ising Magnet Dy$_3$Mg$_2$Sb$_3$O$_{14}$} \\}
\author{Joseph A.~M. Paddison$^{1,2,\dagger}$, Harapan S. Ong$^1$, James O. Hamp$^1$, \\ 
Paromita Mukherjee$^1$, Xiaojian Bai$^2$, Matthew G. Tucker$^{3}$, Nicholas P. Butch$^4$, \\
Claudio Castelnovo$^1$, Martin Mourigal$^2$, and S.~E. Dutton$^{1,\ast}$ \\ \\
	  \normalsize{$^1$Cavendish Laboratory, University of Cambridge, JJ Thomson Avenue, Cambridge, CB3 0HE, UK } \\
	\normalsize{$^2$School of Physics, Georgia Institute of Technology, 	Atlanta, GA 30332, USA }\\
		\normalsize{$^3$ISIS Neutron and Muon Source, Rutherford Appleton Laboratory,} \\
		\normalsize{Harwell Campus, Didcot, OX11 0QX, UK }\\
		\normalsize{$^4$NIST Center for Neutron Research, National Institute of Standards and Technology,} \\ 
		\normalsize{Gaithersburg, MD 20899, USA} \\
		\\
		\normalsize{$^\dagger$Email: paddison@gatech.edu} \\
		\normalsize{$^\ast$Email: sed33@cam.ac.uk} \\
}
\date{April 29, 2016}
\maketitle	
\tableofcontents
\clearpage
\baselineskip18pt


\section{Crystal-structure refinements}\label{sec:data_analysis}

\begin{figure}[h]
\begin{center}
	\includegraphics[scale=0.75] {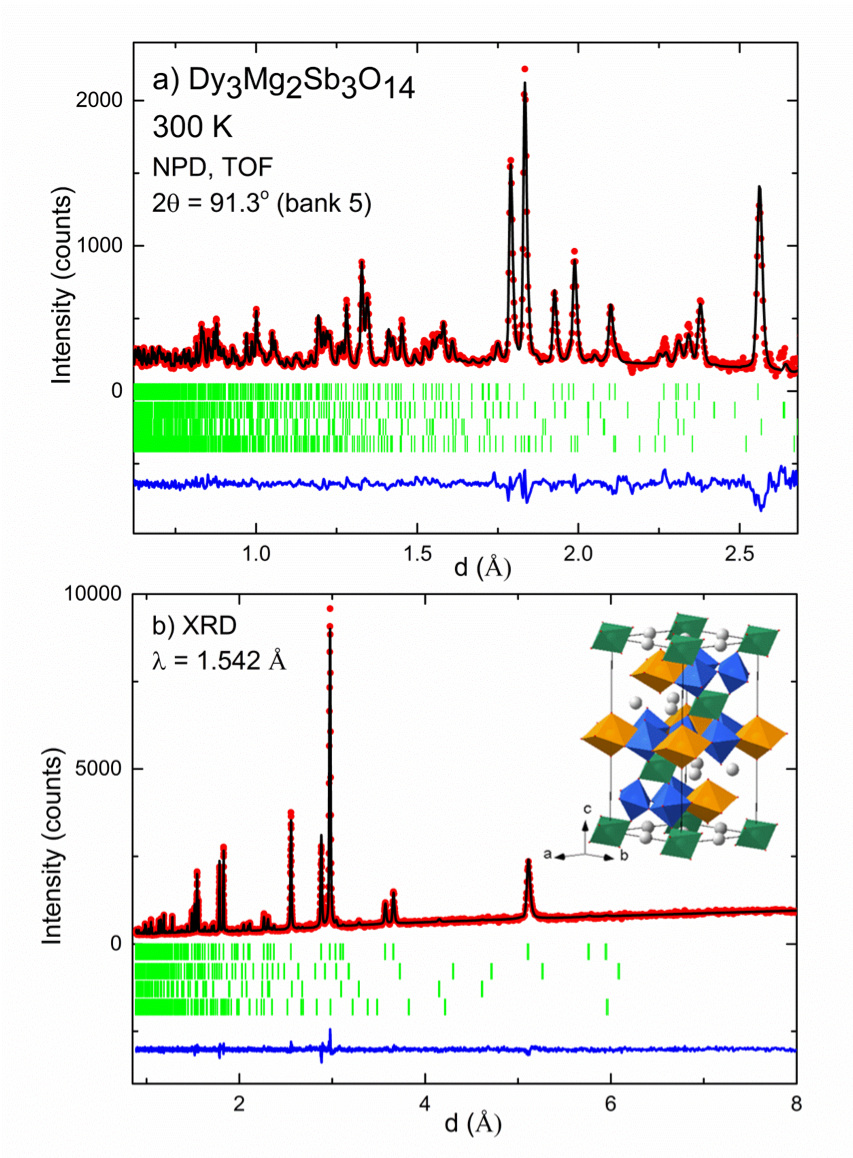}
    \caption{\label{fig:structure1 }Neutron (a) and X-ray (b) diffraction measurements for \dmso~at 300\,K. Observed intensities and calculated intensities obtained from a combined multi-bank Rietveld refinement are shown as red circles and a black line, respectively; the difference is shown by the blue line. Reflection positions are indicated by green tick marks, for phases (top to bottom): \dmso, Dy$_3$SbO$_7$, MgSb$_2$O$_6$, and VO$_2$. A polyhedral model for the structure of \dmso~is inset in (b). Mg1, Mg2(Dy) and Sb1 polyhedra are shown in green, orange, and blue, respectively; Dy1(Mg) sites are shown as white spheres.}
\end{center}
\end{figure}\

\begin{table}[h]
  \begin{centering}
  \begin{tabular}{ccc|cc}
  \hline 
  \multicolumn{5}{c}{\textbf{Dy$_{3}$Mg$_{2}$Sb$_{3}$O$_{14}$ nuclear, $R\bar{3}m$,
  $Z=3$}}\tabularnewline
  \hline 
  \hline 
   &  & $T$ (K) & 300 & 25\tabularnewline
   &  & \multirow{2}{*}{Radiation} & X-ray (Cu $K_{\alpha}$) & Neutron (TOF)\tabularnewline
   &  &  & $+$ Neutron (TOF) & \tabularnewline
  \hline 
   &  & $a$ (\AA) & $7.3217(2)$ & $7.333(10)$\tabularnewline
   &  & $c$ (\AA) & $17.298(2)$ & $17.31(2)$\tabularnewline
   &  & $B_{\mathrm{ov}}$ (\AA$^2$) & $0.2(2)$ & $0^{\ast}$\tabularnewline
  \hline 
  Neutron, Bank 1 & $2\theta=9.39^{\circ}$ & \multirow{6}{*}{$R_{\mathrm{wp}}$} & $7.58$ & $7.16$\tabularnewline
  Neutron, Bank 2 & $2\theta=17.98^{\circ}$ &  & $5.17$ & $5.39$\tabularnewline
  Neutron, Bank 3 & $2\theta=34.96^{\circ}$ &  & $5.49$ & $5.85$\tabularnewline
  Neutron, Bank 4 & $2\theta=63.62{}^{\circ}$ &  & $5.91$ & $5.90$\tabularnewline
  Neutron, Bank 5 & $2\theta=91.30^{\circ}$ &  & $9.65$ & $8.84$\tabularnewline
  X-ray & $\lambda=1.542$\,\AA &  & $5.65$ & $-$\tabularnewline
  \hline 
  \multirow{2}{*}{Dy1} & \multirow{2}{*}{$9e$, $(\frac{1}{2},0,0)$} & Frac. Dy & $0.94(2)$ & $0.94^{\ast}$\tabularnewline
   &  & Frac. Mg & $0.06(2)$ & $0.06^{\ast}$\tabularnewline
  \hline 
  Mg1 & $3a$, $(0,0,0)$ &  &  & \tabularnewline
  \hline 
  \multirow{2}{*}{Mg2} & \multirow{2}{*}{$3b$, $(0,0,\frac{1}{2})$} & Frac. Dy & $0.18(6)$ & $0.18^{\ast}$\tabularnewline
   &  & Frac. Mg & $0.82(6)$ & $0.82^{\ast}$\tabularnewline
  \hline 
  Sb1 & $9d$, $(\frac{1}{2},0,\frac{1}{2})$ &  &  & \tabularnewline
  \hline 
  O1 & $6c$, $(0,0,z)$ & $z$ & $0.384(4)$ & $0.383(4)$\tabularnewline
  \hline 
  \multirow{2}{*}{O2} & \multirow{2}{*}{$18h$, $(x,-x,z)$} & $x$ & $0.531(3)$ & $0.531(2)$\tabularnewline
   &  & $z$ & $0.144(2)$ & $0.144(2)$\tabularnewline
  \hline 
  \multirow{2}{*}{O3} & \multirow{2}{*}{$18h$, $(x,-x,z)$} & $x$ & $0.145(3)$ & $0.145(2)$\tabularnewline
   &  & $z$ & $-0.055(2)$ & $-0.055(2)$\tabularnewline
  \hline 
  \end{tabular}
  \par\end{centering}
  \protect\caption{ \label{tab:refinement}
  Values of refined structural parameters for \dmso~determined from combined analysis of 300\,K X-ray and neutron powder diffraction data, and 25\,K neutron powder diffraction data.  Fixed parameters are denoted by an asterisk ($^{\ast}$).}
\end{table}

\begin{table}[h]
  \begin{centering}
  \begin{tabular}{c|ll}
  \hline 
  \multicolumn{3}{c}{\textbf{Dy$_{3}$Mg$_{2}$Sb$_{3}$O$_{14}$ }}\tabularnewline
  $T$ (K) & $300$ & $25$\tabularnewline
  \hline 
  \hline 
  Dy1--O1 (\AA) & $2.29(3)\times2$ & $2.28(3)\times2$\tabularnewline
  Dy1--O2 (\AA) & $2.53(3)\times2$ & $2.53(3)\times2$\tabularnewline
  Dy1--O3 (\AA) & $2.46(2)\times4$ & $2.46(2)\times4$\tabularnewline
  $\langle$Dy1--O$\rangle$ (\AA) & $2.44$ & $2.43$\tabularnewline
  \hline 
  Mg1--O3 (\AA) & $2.07(2)\times6$ & $2.07(2)\times6$\tabularnewline
  \hline 
  Mg1--O1 (\AA) & $2.01(7)\times2$ & $2.02(7)\times2$\tabularnewline
  Mg2--O2 (\AA) & $2.54(2)\times6$ & $2.538(13)\times6$\tabularnewline
  $\langle$Mg2--O$\rangle$ () & $2.41$ & $2.41$\tabularnewline
  \hline 
  Sb1--O2 (\AA) & $1.98(2)\times4$ & $1.987(13)\times4$\tabularnewline
  Sb1--O3 (\AA) & $1.95(3)\times2$ & $1.95(3)\times2$\tabularnewline
  $\langle$Sb1--O$\rangle$ (\AA) & $1.97$ & $1.97$\tabularnewline
  \hline 
  Dy1--Dy1 (intra-plane) (\AA) & $3.6609(3)\times4$ & $3.663(5)\times4$\tabularnewline
  Dy1--Dy1 (inter-plane) (\AA) & $7.1498(6)\times6$ & $7.157(6)\times6$\tabularnewline
  \hline 
  Dy1--Mg2 (\AA) & $3.5749(3)\times6$ & $3.579(3)\times6$\tabularnewline
  \hline 
  \end{tabular}
  \par\end{centering}
  \protect\caption{\label{tab:angles}Selected bond lengths in \dmso~determined from combined analysis of 300\,K X-ray and neutron powder diffraction data, and 25\,K neutron powder diffraction data.}
\end{table}

\clearpage
\section{Inelastic neutron scattering}

\begin{figure}[h]
\begin{center}
	\includegraphics{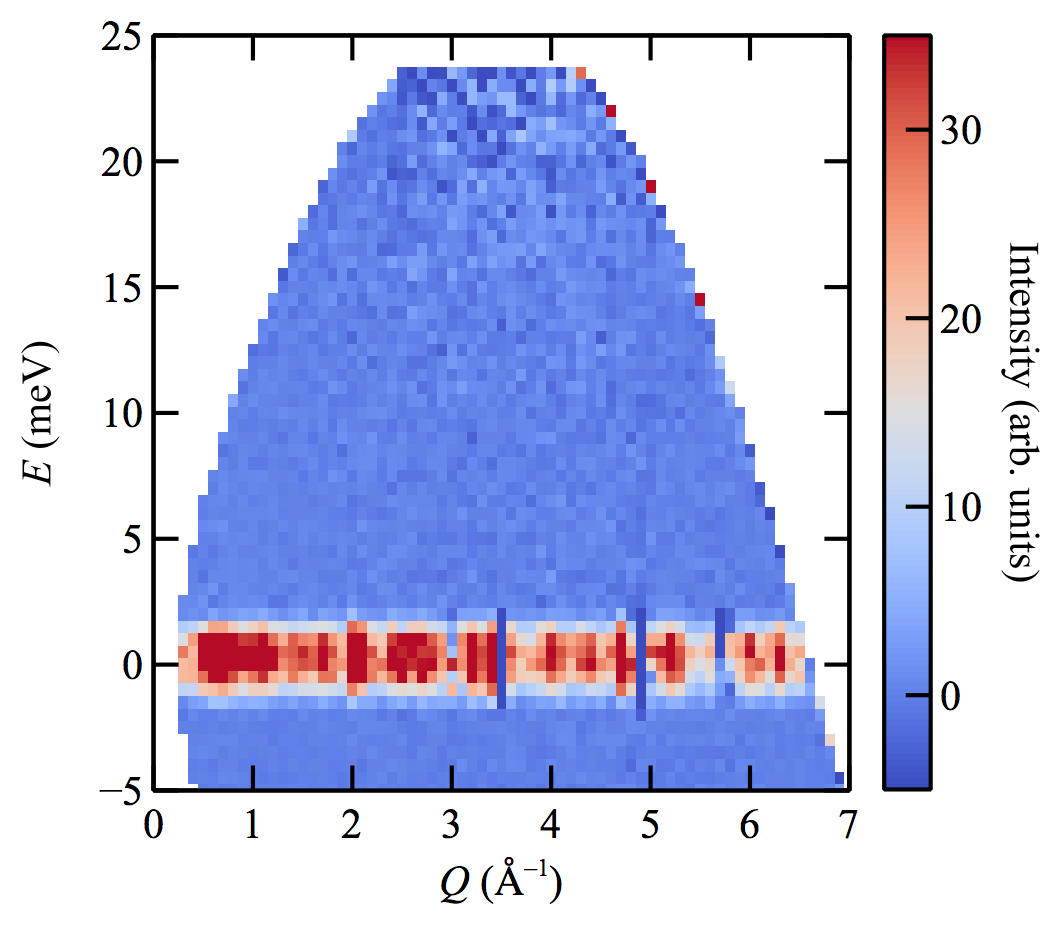}
	\caption{\label{fig:cef}
	Momentum and energy dependence of neutron-scattering intensity at nominal $T=0.03$\,K, measured using the DCS spectrometer at NIST \cite{Copley_2003} with incident neutron wavelength $\lambda=1.8$\,\AA. An empty can measurement has been subtracted from the data in order to remove background scattering. These measurements reveal no crystal electric-field excitations to a maximum energy transfer 23\,meV.}
\end{center}
\end{figure}

\clearpage
\section{Heat capacity}\label{sub:analysis_heat_capacity}

\begin{figure}[h]
\begin{center}
	\includegraphics[scale=0.83]{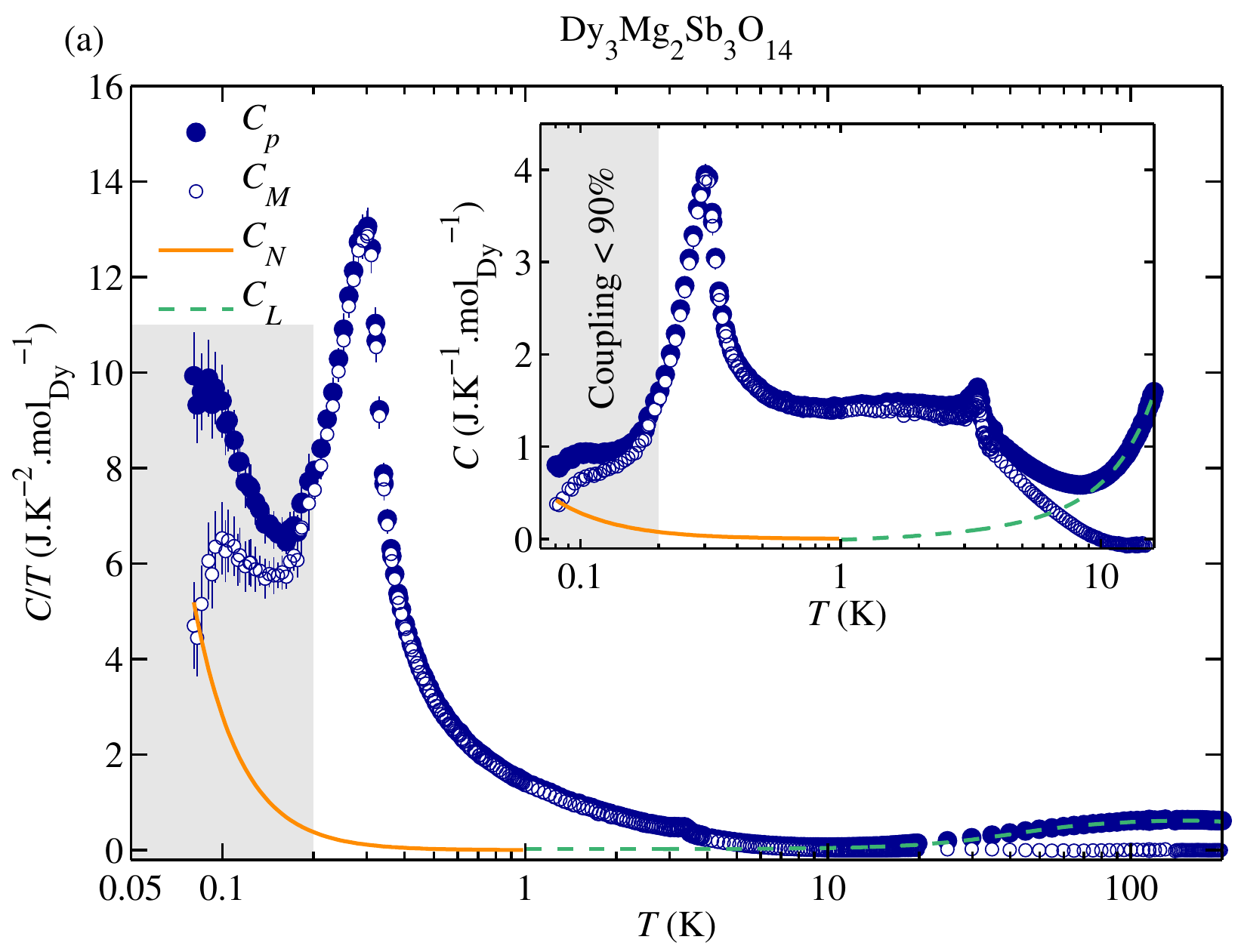} 
    \caption{\label{fig:hcmeas} Specific heat of  \dmso, showing the following contributions: total measured ($C_p$, full blue circles), modelled lattice ($C_\mathrm{l} $, green dashed line), modelled nuclear ($C_\mathrm{n}$, solid orange line), and extracted magnetic ($C_\mathrm{m}=C_p  - C_\mathrm{n} - C_\mathrm{l}$, open blue circles). The main panel shows $C/T$ and the inset shows $C$. The temperature region where sample coupling falls below 90\% is shaded grey. The magnetic specific heat $C_\mathrm{m}$ displays a very small peak at $T_i=3.35(5)$\,K, which is probably associated with the magnetic ordering of the $\sim$\,2\,wt\% Dy$_3$SbO$_7$ impurity phase~\cite{Fennell_2001}. There is also a small peak in $C_\mathrm{m}$ around 0.1\,K, which is of uncertain origin given the poor sample coupling below approximately $0.2$\,K.}
\end{center}
\end{figure}

\clearpage
\section{Quasi-static approximation}\label{sub:quasistatic approximation}

\begin{figure}[h]
\begin{center}
	\includegraphics{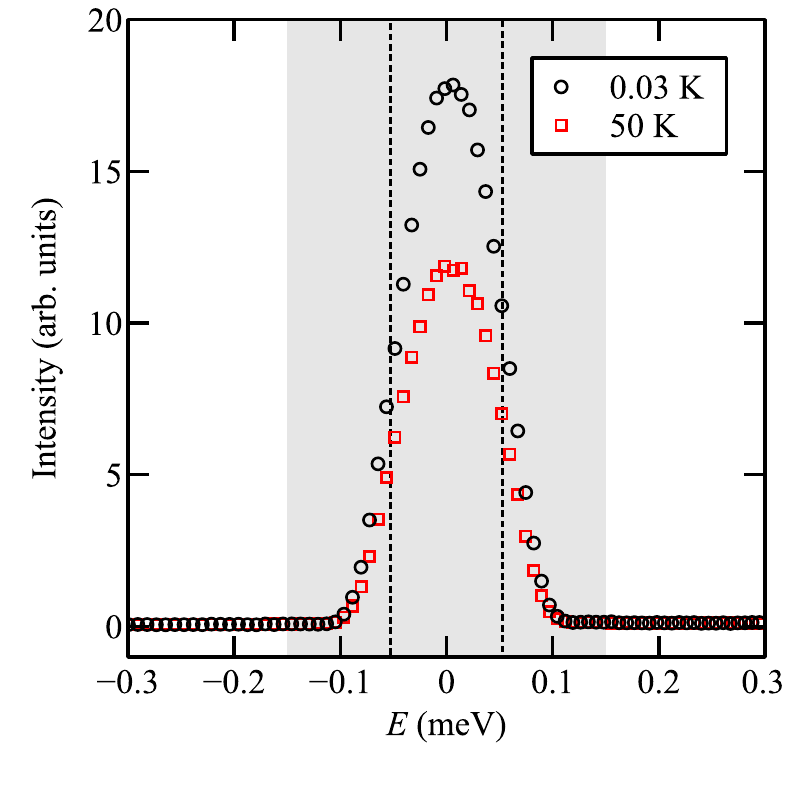} 
	\caption{\label{fig:quasistatic} Energy dependence of neutron-scattering intensity at 0.03\,K
	(black circles) and 50\,K (red squares), measured with incident neutron wavelength $\lambda=5$\,\AA~using the DCS spectrometer at NIST \cite{Copley_2003}. The data are integrated over
	the range $15\leq2\theta\leq45^{\circ}$, which contains no nuclear
	Bragg peaks but intense magnetic diffuse scattering.
	The black dotted vertical lines show the calculated energy resolution
	of the instrument, and the grey box
	shows the range of energy integration for the neutron-scattering
	data shown in Fig.~2b. The energy line-width due to dynamical spin fluctuations is limited by the instrumental resolution function at both 0.03 and $50.0$\,K, which demonstrates that the quasi-static approximation is ideally satisfied. The decrease in peak intensity at 50\,K occurs because diffuse intensity is redistributed to higher $2\theta$.}
\end{center}
\end{figure}

\clearpage
\section{Magnetic structure of impurity Dy$_3$SbO$_7$}\label{sub:analysis_impurity}

Magnetic Bragg peaks appear in our neutron-diffraction data at $T\leq2$\,K, but are absent in $T\geq4$\,K data. These peaks cannot be indexed by high-symmetry propagation vectors of \dmso, but are indexed by the propagation vector $\mathbf{k}=(0,0,0)$ for the Dy$_3$SbO$_7$ impurity phase \cite{Siqueira_2013}. We therefore identify these peaks with magnetic ordering of Dy$_3$SbO$_7$, which is reported to occur at $3.0(3)$\,K \cite{Fennell_2001}.

\begin{figure}[h]
\begin{center}
	\includegraphics{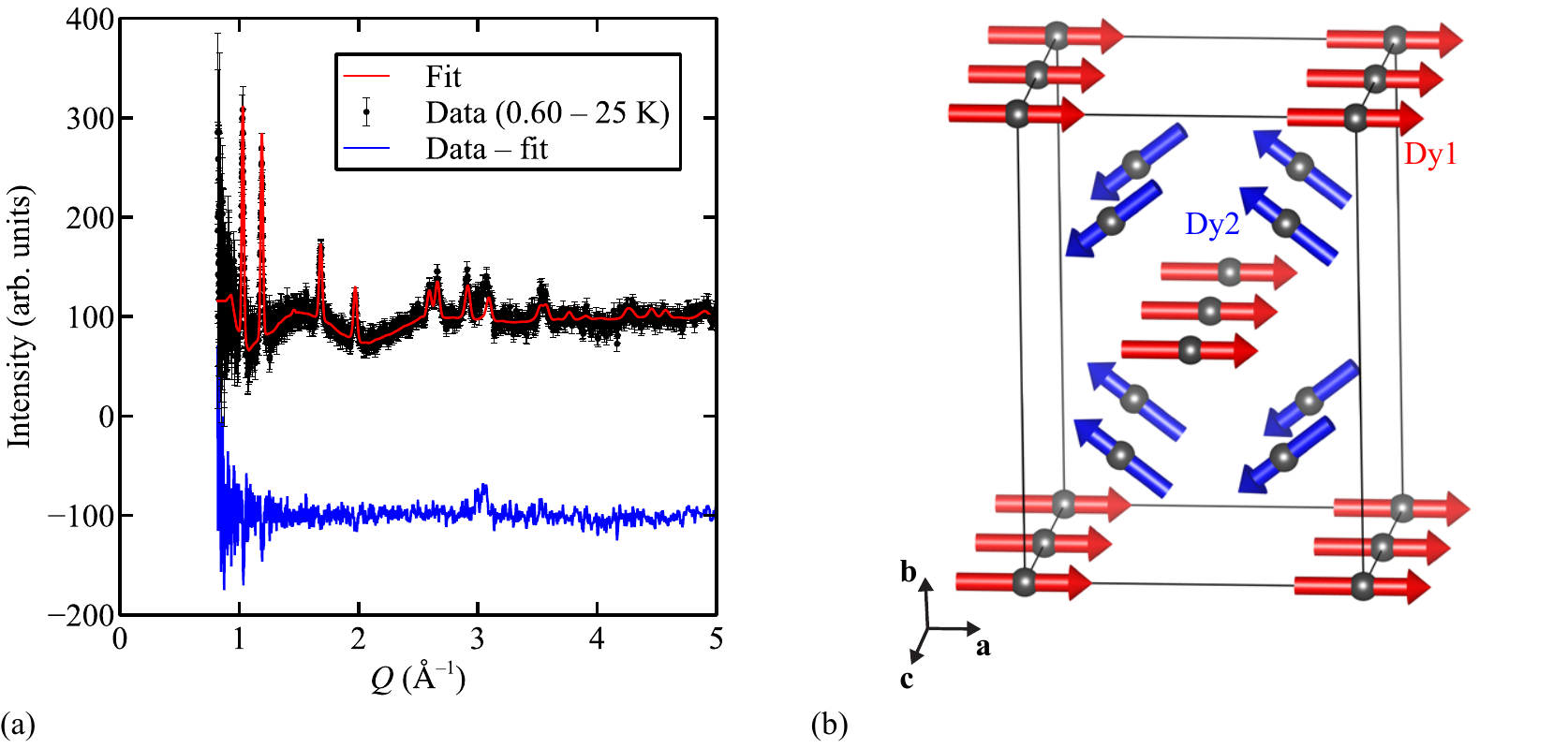}
	\caption{\label{fig:mag_Dy3SbO7}
	(a) Rietveld fits to the magnetic Bragg scattering
	data for Dy$_3$SbO$_7$ obtained by subtracting 25\,K from $0.60$\,K GEM data, as described in the text. Experimental data
	are shown as black circles, Rietveld fit as a red line, and difference
	(data--fit) as a blue line. (b) Model of the magnetic structure of Dy$_3$SbO$_7$ obtained from Rietveld refinement. Orientations of ordered magnetic moments are shown as red arrows (Dy1 site) and blue arrows (Dy2 site).}
\end{center}
\end{figure}

\begin{table}
	\centering{}%
	\begin{tabular}{ccc|ccc}
	\hline 
	Site & Basis vector, $\nu$  & Atom, $i$ & $m_{a}$ & $m_{b}$ & $m_{c}$\tabularnewline
	\hline 
	\hline 
	Dy1 & 1 & 1 & $4$ & $0$ & $0$\tabularnewline
	 &  & 2 & $4$ & $0$ & $0$\tabularnewline
		\hline 
	Dy2 & 1 & 3 & $2$ & $0$ & $0$\tabularnewline
	 &  & 4 & $2$ & $0$ & $0$\tabularnewline
	 &  & 5 & $2$ & $0$ & $0$\tabularnewline
	 &  & 6 & $2$ & $0$ & $0$\tabularnewline
	 & 2& 3 & $0$ & $2$ & $0$\tabularnewline
	 &  & 4 & $0$ & $-2$ & $0$\tabularnewline
	 &  & 5 & $0$ & $-2$ & $0$\tabularnewline
	 &  & 6 & $0$ & $2$ & $0$\tabularnewline
	\hline 
	\end{tabular}\protect\caption{\label{tab:mag_basis_Dy3SbO7} Basis vectors for the atoms at fractional
	coordinates $\mathbf{r}_{1}=(0,0,0)$, $\mathbf{r}_{2}=(0,0,\frac{1}{2})$,
	$\mathbf{r}_{3}=(x,y,\frac{3}{4})$, $\mathbf{r}_{4}=(x,-y,\frac{1}{4})$, $\mathbf{r}_{5}=(-x,y,\frac{3}{4})$, and $\mathbf{r}_{6}=(-x,-y,\frac{1}{4})$
	in the orthorhombic unit cell of Dy$_3$SbO$_7$.}
\end{table}

\begin{table}
  \begin{centering}
  \begin{tabular}{ccc|c}
  \hline 
  \multicolumn{4}{c}{\textbf{Dy$_{3}$SbO$_{7}$ magnetic, $Cmcm$}}\tabularnewline
  \hline 
  \hline 
   &  & $T$ (K) & $0.6-25$\tabularnewline
   &  & \multirow{1}{*}{Radiation} & Neutron (TOF)\tabularnewline
  \hline 
   &  & $a$ (\AA) & $7.451615(7)$\tabularnewline
   &  & $b$ (\AA) & $10.533144(3)$\tabularnewline
   &  & $c$ (\AA) & $7.446(3)$\tabularnewline
  \hline 
  Bank 1 & $2\theta=9.39^{\circ}$ & \multirow{4}{*}{$R_{\mathrm{wp}}$} & $11.6$\tabularnewline
  Bank 2 & $2\theta=17.98^{\circ}$ &  & $10.1$\tabularnewline
  Bank 3 & $2\theta=34.96^{\circ}$ &  & $7.25$\tabularnewline
  Bank 4 & $2\theta=63.62{}^{\circ}$ &  & $7.26$\tabularnewline
  \hline 
  Dy1 & $4a$, $(0,0,0)$ & $C_{1}$ & $2.7(3)$\tabularnewline
  \hline 
  \multirow{4}{*}{Dy2} & \multirow{4}{*}{$8g$, $(x,y,\frac{3}{4})$} & $x$ & $0.259(21)$\tabularnewline
   &  & $y$ & $0.231(4)$\tabularnewline
   &  & $C_{1}$ & $-3.7(3)$\tabularnewline
   &  & $C_{2}$ & $2.9(3)$\tabularnewline
  \hline 
  \end{tabular}\protect\caption{\label{tab:mag_params_Dy3SbO7} Refined values of structural parameters for Dy$_3$SbO$_7$ for the magnetic-structure model described in the text. The basis vector coefficients $C_\nu$ are determined up to an overall scale factor.}
  \par\end{centering}
  \end{table}

A magnetic-structure model for Dy$_3$SbO$_7$ has not been reported previously, so we turn to symmetry analysis to identify possible magnetic structures. The crystal structure of Dy$_3$SbO$_7$ (space group $Cmcm$ \cite{Siqueira_2013}) contains two inequivalent Dy sites, $4a$ (Dy1) and $8g$ (Dy2). The symmetry-allowed magnetic structures were determined by representational analysis using the program Sarah \cite{Wills_2000}. The analysis first determines the group of symmetry elements that leave $\mathbf{k}$ invariant, and then decomposes the magnetic representations of Dy1 and Dy2 sites into irreducible representations (irreps, $\Gamma$) of this group.The decomposition of the magnetic representation for the Dy1 site is 
\begin{equation}
	\Gamma_{\mathrm{Mag}}\left(\mathrm{Dy1}\right)=1\Gamma_{1}^{1}+0\Gamma_{2}^{1}+1\Gamma_{3}^{1}+0\Gamma_{4}^{1}+2\Gamma_{5}^{1}+0\Gamma_{6}^{1}+2\Gamma_{7}^{1}+0\Gamma_{8}^{1},\label{eq:neutron_Dy1}
	\end{equation}
	 and the decomposition for the Dy2
	site is 
	\begin{equation}
	\Gamma_{\mathrm{Mag}}\left(\mathrm{Dy2}\right)=1\Gamma_{1}^{1}+2\Gamma_{2}^{1}+2\Gamma_{3}^{1}+1\Gamma_{4}^{1}+2\Gamma_{5}^{1}+1\Gamma_{6}^{1}+1\Gamma_{7}^{1}+2\Gamma_{8}^{1},\label{eq:neutron_Dy2}
\end{equation}
where different irreps are labelled by subscript numbers (using the notation of Kovalev \cite{Kovalev_1993}), the dimensionality of each irrep is given by the superscript number, and the product of the superscript and the coefficient yields the number of basis vectors associated with the irrep. Because a single magnetic transition is reported in Dy$_3$SbO$_7$ \cite{Fennell_2001}, we consider only the irreps that appear in the decomposition for both Dy1 and Dy2 sites---namely, $\Gamma_1,\Gamma_3,\Gamma_5$, and $\Gamma_7$. We used Rietveld refinement to test each of these irreps against the $0.60-25$\,K data collected on the GEM diffractometer at ISIS \cite{Hannon_2005}. In these refinements, the magnetic peak-shape parameters were modelled as convoluted pseudo-Voigt and Ikeda-Carpenter functions, and the background was fitted by a linear interpolation between between fixed points (for which the background level was subsequently refined at lower $Q$). The $\Gamma_{3}$ irrep provided the best fit-to-data, shown in Fig.~S\ref{fig:mag_Dy3SbO7}a. This magnetic structure is shown in Fig.~S\ref{fig:mag_Dy3SbO7}b and describes a non-collinear ferrimagnet. Because of the uncertainty associated with the wt\% Dy$_3$SbO$_7$ in our sample, it was not possible to determine accurately the values of the ordered magnetic moment on Dy1 and Dy2 sites; however, the data are consistent with moment lengths on Dy1 and Dy2 sites that are equal to within $\sim$20\%. The orientation of the ordered moment of atom $i$ is given by
\begin{equation}
	\boldsymbol{\mu}_{i,\mathrm{avg}}=\sum_{\nu}C_{\nu}\mathbf{m}_{\nu,i},
	\end{equation}
	where the $C_{v}$ denotes the refined coefficient of the basis vector
	\begin{equation}
	\mathbf{m}_{\nu,i}=m_{a}^{\nu,i}\hat{\mathbf{a}}+m_{b}^{\nu,i}\hat{\mathbf{b}}+m_{c}^{\nu,i}\hat{\mathbf{c}},
\end{equation}
where $\hat{\mathbf{a}},\hat{\mathbf{b}},\hat{\mathbf{c}}$ are unit
vectors parallel to the unit-cell axes. The projections $m_{a},m_{b},m_{c}$
are given in Table~S\ref{tab:mag_basis_Dy3SbO7}, and refined values of magnetic-structure parameters are given in Table~S\ref{tab:mag_params_Dy3SbO7}.

Having determined a model of the magnetic structure of the Dy$_3$SbO$_7$ impurity phase, we calculated the magnetic scattering from \dmso~as $I=I_{\mathrm{meas}}-I_{T_{\mathrm{high}}}-I_{\mathrm{imp,Bragg}}+I_{\mathrm{imp,pm}}$, where $I_{T_{\mathrm{high}}}$ denotes a high-temperature (25\,K or 50\,K) measurement which is subtracted to remove non-magnetic scattering, $I_{\mathrm{imp,Bragg}}$ denotes the calculated magnetic Bragg profile for the Dy$_3$SbO$_7$ impurity, and $I_{\mathrm{imp,pm}}$ denotes the calculated paramagnetic intensity for the Dy$_3$SbO$_7$ impurity.

\clearpage
\section{Average magnetic structure of \dmso}\label{sub:analysis_mag_structure}

Magnetic Bragg peaks from \dmso~are observed at temperatures below $0.5$\,K, but are absent at $0.5$\,K and higher temperatures [Fig.~2b]. To isolate this magnetic Bragg scattering, we subtracted the $0.5$\,K data from the $0.03$, $0.10$, $0.20$, $0.30$ and $0.35$\,K data. All magnetic Bragg peaks are indexed by the propagation vector $\mathbf{k}=\left(0,0,0\right)$ with respect to the hexagonal unit cell of \dmso. Symmetry-allowed magnetic structures were determined using the program Sarah \cite{Wills_2000} and verified using Isodistort \cite{Stokes_2006,Campbell_2006}. The decomposition of the magnetic representation for the Dy1 site is 
\begin{equation}
	\Gamma_{\mathrm{Mag}}\left(\mathrm{Dy1}\right)=1\Gamma_{1}^{(1)}+0\Gamma_{2}^{(1)}+2\Gamma_{3}^{(1)}+0\Gamma_{4}^{(1)}+2\Gamma_{5}^{(2)}+0\Gamma_{6}^{(2)},\label{eq:neutron_magMg}
	\end{equation}
	 and the decomposition for the Mg2
	site is 
	\begin{equation}
	\Gamma_{\mathrm{Mag}}\left(\mathrm{Mg2}\right)=0\Gamma_{1}^{(1)}+0\Gamma_{2}^{(1)}+1\Gamma_{3}^{(1)}+0\Gamma_{4}^{(1)}+1\Gamma_{5}^{(2)}+0\Gamma_{6}^{(2)}.\label{eq:neutron_magDy}
\end{equation}

\begin{figure}[h]
\begin{center}
	\includegraphics{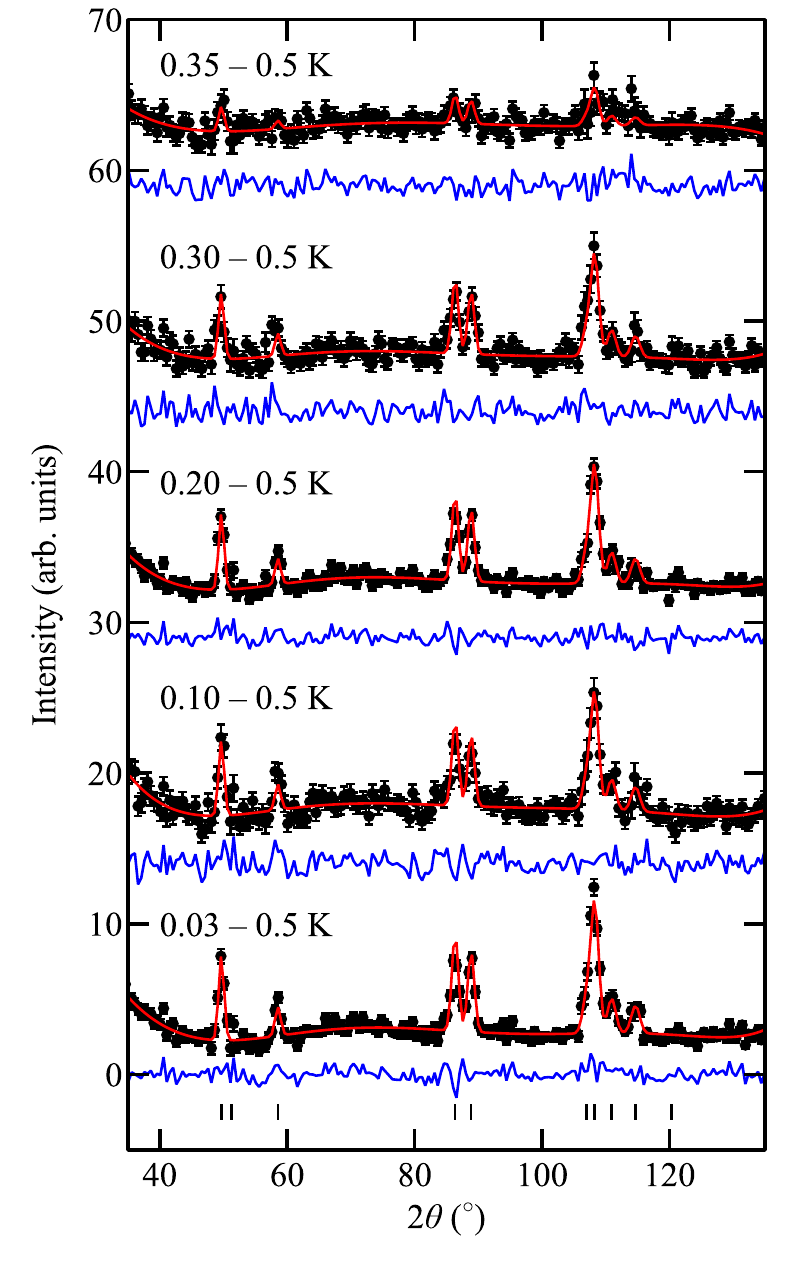}
	\caption{\label{fig:mag_rietveld}
	Rietveld fits to the magnetic Bragg scattering
	data for \dmso~obtained by subtracting 0.5\,K data from the the $0.10$, $0.20$,
	$0.30$ and $0.35$\,K DCS data. Temperatures
	are labelled on the graph. At each temperature, experimental data
	are shown as black circles, Rietveld fits as a red line, and difference
	(data\textendash fit) as a blue line.}
\end{center}
\end{figure}

\begin{table}[h]
	\centering{}%
	\begin{tabular}{ccc|ccc}
	\hline 
	Site & Basis vector, $\nu$  & Atom, $i$ & $m_{a}$ & $m_{b}$ & $m_{c}$\tabularnewline
	\hline 
	\hline 
	Dy1 & 1 & 1 & $2$ & $4$ & $4$\tabularnewline
	 &  & 2 & $-4$ & $-2$ & $4$\tabularnewline
	 &  & 3 & $2$ & $-2$ & $4$\tabularnewline
	 & 2 & 1 & $-2$ & $-4$ & $0$\tabularnewline
	 &  & 2 & $4$ & $2$ & $0$\tabularnewline
	 &  & 3 & $-2$ & $2$ & $0$\tabularnewline
		\hline 
	Mg2 & 1 & 4 & 0 & 0 & 12\tabularnewline
	\hline 
	\end{tabular}\protect\caption{\label{tab:mag_basis_vectors} Basis vectors for the atoms at fractional
	coordinates $\mathbf{r}_{1}=(\frac{1}{2},0,0)$, $\mathbf{r}_{2}=(0,\frac{1}{2},0)$,
	$\mathbf{r}_{3}=(\frac{1}{2},\frac{1}{2},0)$, and $\mathbf{r}_{4}=(0,0,\frac{1}{2})$
	in the hexagonal unit cell of \dmso.}
\end{table}

\begin{table}[h]
  \begin{centering}
  \begin{tabular}{ccc|cc}
  \hline 
  \multicolumn{5}{c}{\textbf{Dy$_{3}$Mg$_{2}$Sb$_{3}$O$_{14}$ magnetic, $R\bar{3}m$}}\tabularnewline
   &  &  & \textbf{2-site model} & \textbf{1-site model}\tabularnewline
  \hline 
  \hline 
   &  & $T$ (K) & \multicolumn{2}{c}{$0.04-0.50$}\tabularnewline
   &  & \multirow{1}{*}{Radiation} & \multicolumn{2}{c}{Neutron ($\lambda=5.0$\,\AA)}\tabularnewline
  \hline 
   &  & $a$ (\AA) & $7.298(3)$ & $7.295(3)$\tabularnewline
   &  & $c$ (\AA) & $17.279(6)$ & $17.277(6)$\tabularnewline
  \hline 
   &  & \multirow{1}{*}{$R_{\mathrm{wp}}$} & $7.54$ & $7.63$\tabularnewline
  \hline 
  \multirow{3}{*}{Dy1} & \multirow{3}{*}{$9e$, $(\frac{1}{2},0,0)$} & $C_{1}$ & $-0.30(2)$ & $-0.31(2)$\tabularnewline
   &  & $C_{2}$ & $-1.03(2)$ & $-1.04(2)$\tabularnewline
   &  & $\mu_{\mathrm{avg}}$ ($\mu_{\mathrm{B}}$ per Dy) & $2.80(4)$ & $2.82(4)$\tabularnewline
  \hline 
  \multirow{2}{*}{Mg2} & \multirow{2}{*}{$3b$, $(0,0,\frac{1}{2})$} & $C_{1}$ & $-0.08(4)$ & $0^{\ast}$\tabularnewline
   &  & $\mu_{\mathrm{avg}}$ ($\mu_{\mathrm{B}}$ per Dy) & $1.0(5)$ & $0^{\ast}$\tabularnewline
  \hline 
  \end{tabular}
  \par\end{centering}
  \protect\caption{\label{tab:mag_rietveld_params} Values of structural parameters 
  	obtained for the two magnetic-structure models of \dmso~described in the text. Fixed parameters are denoted by an asterisk ($^{\ast}$).}
\end{table}

Rietveld refinements were performed to the $0.03-0.50$\,K data to determine the average magnetic structure. In these refinements, the occupancy of the Dy site was fixed at 94\%, as estimated from the crystal-structure refinements [Section~\ref{sec:data_analysis}]. The magnetic peak-shape parameters were modelled as Gaussian and fixed to equal the nuclear peak-shape parameters refined to 50\,K data. The intensity scale factor was fixed at the value obtained from Rietveld refinement to the 50\,K data, and the background was fitted by Chebychev polynomials. By testing each of the irreps $\Gamma_{1}$, $\Gamma_{3}$, and $\Gamma_{5}$ against the data, we found that only the $\Gamma_{3}$ irrep allowed a good fit. The basis vectors for this magnetic structure are shown in Table~S\ref{tab:mag_basis_vectors}. The component of the ordered magnetic moment parallel to the $c$-axis on the Dy1 site, $\mu_{c,\mathrm{avg}}/\mu_{\mathrm{avg}}=0.44(3)$, shows that Dy spins are canted slightly more towards the $c$-axis than for cubic spin ices such as Dy$_{2}$Ti$_{2}$O$_{7}$, where the equivalent projection equals $1/3$. 

Because $\Gamma_{3}$ occurs in the magnetic representation for both Dy1 and Mg2 sites, ordered moments may form on both sites in a single phase transition. The basis vectors of the Dy1 site describes an ``all-in/all-out'' structure, while basis vectors of the Mg2 site describe a uniform ferromagnetic component along the $c$-axis. Mindful of the fact that weak ferromagnetic components can be poorly-determined from powder-diffraction data, we performed two separate refinements to the $0.03-0.50$\,K data. First, we refine the basis-vector coefficients on both sites (``2-site refinement''); second, we constrain the magnetic moment on the Mg2 site to equal zero (``1-site refinement''). Table~S\ref{tab:mag_rietveld_params} shows the results from each refinement. Our data are consistent with a small ordered moment ($\sim $\,$1 \mu_{\mathrm{B}}$) on the Mg2 site; however, refining this moment yields an insignificant improvement in the fit, and does not change parameter values associated with the Dy1 site. In subsequent refinements, we therefore fix $\mu_{\mathrm{avg}}\equiv0$ on the Mg2 site for the sake of simplicity. The fits obtained to $0.10$, $0.20$, $0.30$ and $0.35$\,K data are shown in Fig.~S\ref{fig:mag_rietveld}. For these refinements, we fixed the magnetic structure to the result from the refinement to $0.03-0.50$\,K data, and refined only the background parameters and the overall scale factor in order to determine the temperature dependence of $\mu_\mathrm{avg}$ [Fig.~3a]. 

\clearpage
\section{Reverse Monte Carlo refinements}\label{sec:rmc}

\begin{figure}[hhh]
\begin{center}
	\includegraphics{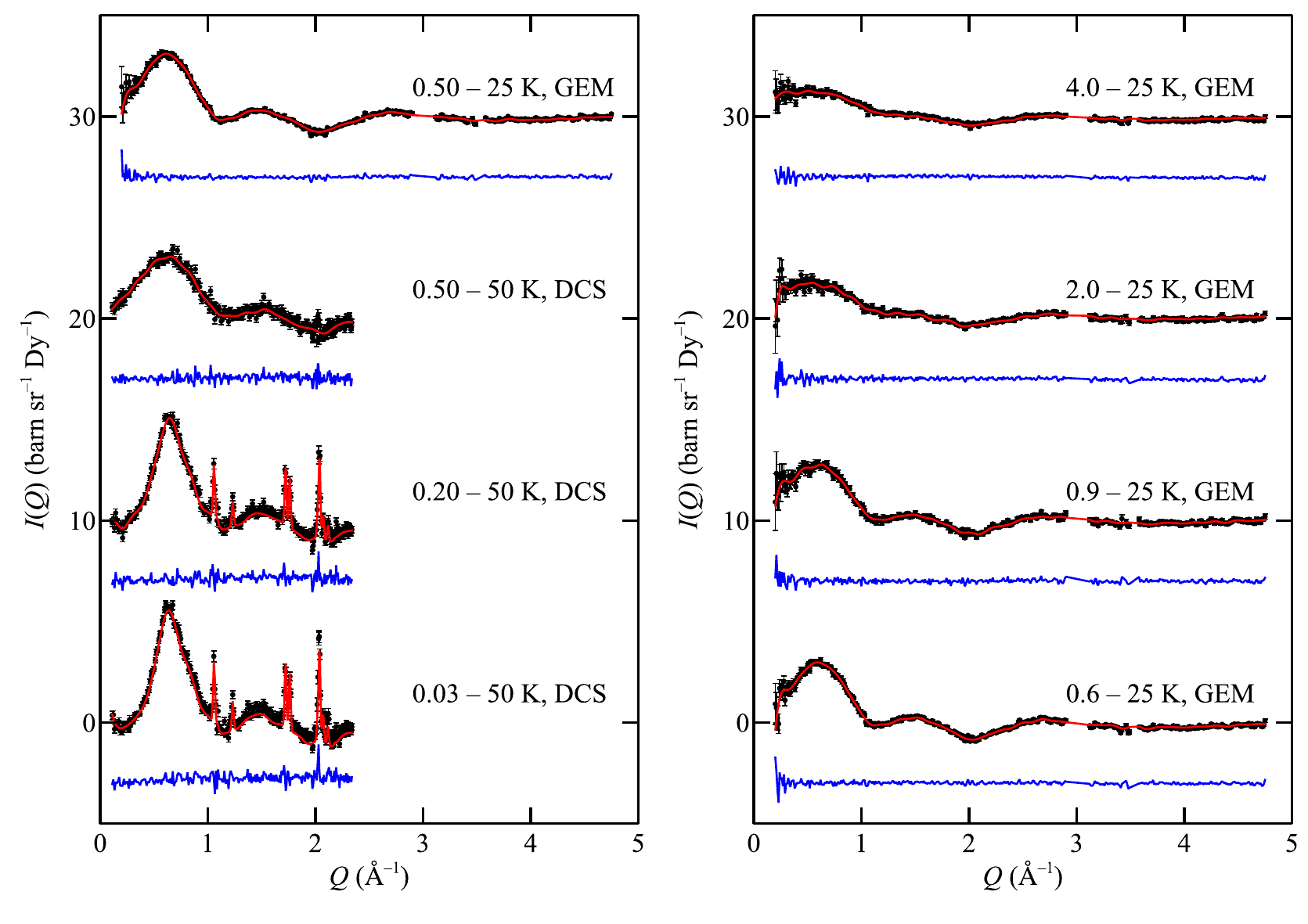}
	\caption{\label{fig:rmc_fits}
	Reverse Monte Carlo fits to neutron-scattering data collected on the DCS and GEM neutron-scattering instruments, showing experimental data (black circles), RMC fits (red lines), and data--fit (blue lines). The temperature and instrument used (DCS or GEM) are labelled above each curve. Successive curves are vertically shifted by $10$~barn$\,$sr$^{-1}\,$Dy$^{-1}$ for clarity.}
\end{center}
\end{figure}

Fig.~S\ref{fig:rmc_fits} shows representative reverse Monte Carlo (RMC) fits to neutron-scattering data collected on DCS and GEM instruments at different temperatures. The resulting spin configurations were used to calculate the \% of triangles with $\pm3$ charge [Fig.~3a], and the charge correlation function [Fig.~3b].

\clearpage
\section{Monte Carlo simulations}\label{sec:mc}

\begin{figure}[h]
\begin{center}
	\includegraphics{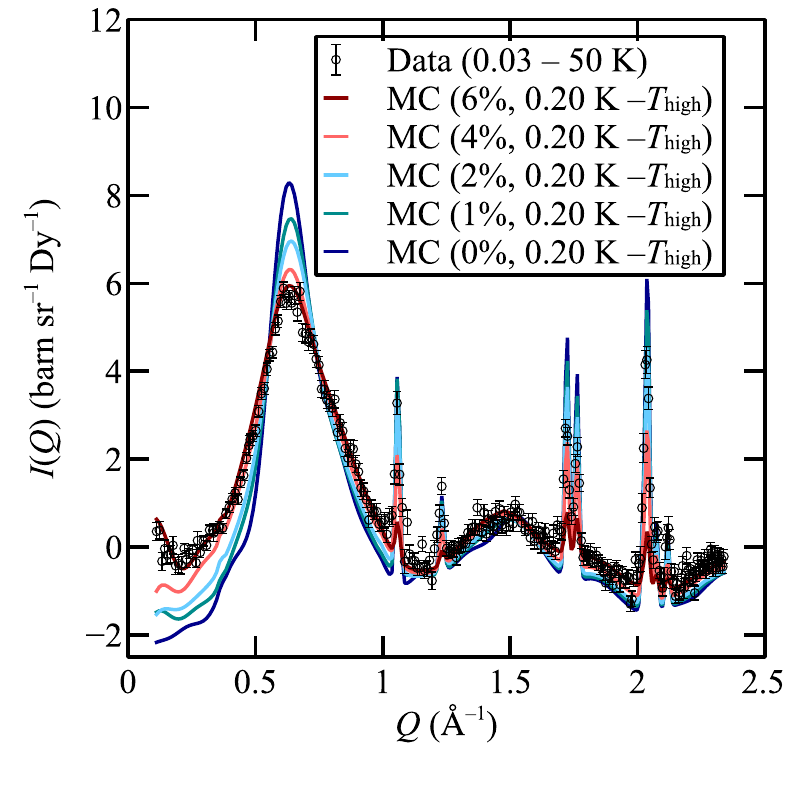}
	\caption{\label{fig:powder_mc}
	Powder-averaged magnetic neutron-scattering patterns calculated from Monte Carlo simulations for different amounts of random site disorder (expressed as \% Mg on Dy1 site). Calculations (solid coloured lines) are shown for 0.2\,K; experimental data collected at 0.2\,K (hollow black circles) are shown for comparison.}
\end{center}
\end{figure}

Fig.~S\ref{fig:powder_mc} shows the powder-averaged magnetic neutron-scattering patterns calculated from Monte Carlo simulations for different amounts of random site disorder. Calculations are performed using the same approach as the RMC refinements (described in the Methods section), and a total magnetic moment $\mu =10.0\,\mu_\mathrm{B}$ per Dy is assumed. No fitting parameters are included in order to match the experimental data. Good overall agreement is achieved between 0.2\,K simulations and experimental data for between $\sim$4 and 6\% Mg on the Dy1 site; values close to 4\% yield best agreement with Bragg intensities, whereas values close to 6\% yield best agreement with the diffuse-scattering profile. We anticipate that the agreement may be further improved by fitting the value of the nearest-neighbour exchange interaction and/or by considering possible short-range correlations of Mg occupancy of the Dy1 site. 

\section*{Supplementary References}
\begingroup
\renewcommand{\section}[2]{}
		 

\begin{thebibliography}{99}
\bibitem{NIST}
The name of a commercial product or trade name does not imply endorsement or recommendation by NIST.
\end{thebibliography}
\end{document}